\documentclass[11pt,epsfig]{article}

\usepackage[reqno]{amsmath}
\usepackage{epsfig}
\usepackage{array}
\usepackage{float}

\def\ltap{\ \raisebox{-.4ex}{\rlap{$\sim$}} \raisebox{.4ex}{$<$}\ }
\def\gtap{\ \raisebox{-.4ex}{\rlap{$\sim$}} \raisebox{.4ex}{$>$}\ }
\def\ltap{\ \raisebox{-.4ex}{\rlap{$\sim$}} \raisebox{.4ex}{$<$}\ }
\def\gtap{\ \raisebox{-.4ex}{\rlap{$\sim$}} \raisebox{.4ex}{$>$}\ }

\begin{document}

\rightline{Ref. SISSA 91/2001/EP}
\rightline{December 2001}
\rightline{hep-ph/0201087}
\vskip 0.6cm
% ~\vfill
\begin{center}
{\bf
On the Day-Night Effect and CC to NC Event 
Rate Ratio Predictions\\
for the SNO Detector\\  
% \footnote{Presented by S. T. Petcov}
}

\vspace{0.3cm} 
M. Maris$^{(a)}$~~and ~~S. T. Petcov$^{(b,c)}$~
\footnote{Also at: Institute of Nuclear Research and
Nuclear Energy, Bulgarian Academy of Sciences, 1784 Sofia, Bulgaria}

\vspace{0.2cm}   
{\em $^{(a)}$ Osservatorio Astronomico di Trieste, I-34113 Trieste, Italy\\
}
 
\vspace{0.2cm}   
{\em $^{(b)}$ Scuola Internazionale Superiore di Studi Avanzati, 
I-34014 Trieste, Italy\\
}
\vspace{0.2cm}   
{\em $^{(b)}$ Istituto Nazionale di Fisica Nucleare, 
Sezione di Trieste, I-34014 Trieste, Italy\\
}

\end{center}
\vskip 0.4cm
\begin{abstract}
Detailed predictions   
for the D-N asymmetry for the Super-Kamiokande
experiment, as well as for the 
{\it Full Night} and {\it Core}
D-N asymmetries in the solar neutrino induced
CC event rate 
% in the SNO detector,
and the {\it Day}, {\it Night} and 
{\it Core} ratios of the 
CC and NC event rates,
measured in the SNO experiment, 
are derived in the cases of the LMA MSW 
and LOW solutions of the solar neutrino problem. 
The indicated observables for the SNO
experiment are calculated
for two values of the 
threshold (effective)
kinetic energy of the final state 
electron in the CC reaction on deuterium:
$T_{\mathrm{e},\mathrm{th}} = 6.75$\ MeV
and 5.0 MeV. The possibilities 
to further constrain the regions of 
the LMA MSW and LOW solutions of the 
solar neutrino problem
by using the forthcoming SNO data
on the D-N asymmetry and on the 
CC to NC event rate ratio are 
also discussed.
\end{abstract}
% \leftline{PACS: 14.60.Pq, 23.40.-s}
\vfill
% \normalsize\baselineskip=15pt
\newpage
\vspace{-0.3cm}
\section{Introduction} 
\vspace{-0.2cm}

\hskip 0.5cm The recent SNO results \cite{SNO1},
combined with the data from the Super-Kamiokande
experiment \cite{SKsol}, clearly demonstrate the 
presence of $\nu_{\mu}$ ( $\nu_{\tau}$) component
in the flux of solar neutrinos reaching the Earth
\footnote{The non-electron neutrino component 
in the flux of solar neutrinos can also
include, or correspond to, $\bar{\nu}_{\mu}$ and/or 
$\bar{\nu}_{\tau}$ \cite{SNO1}.}.
This represents a 
compelling evidence for 
oscillations and/or transitions
of the solar neutrinos.

  The SNO experiment 
measured the rate of the 
charged current (CC) reaction
$\nu_e + D \rightarrow e^{-} + p + p$
for $T_e \geq 6.75~{\rm MeV}$, $T_e$ being the 
(effective) kinetic energy 
of the final state electron \cite{SNO1}.
The reaction is due to the flux of
solar $\nu_e$ from $^8$B decay
having energy of $E \gtap 8.2~{\rm MeV}$.
Assuming that the $^8$B 
neutrino energy spectrum 
is not substantially modified by 
the solar neutrino oscillations, 
the SNO collaboration obtained the 
following value of the solar 
$\nu_e$ flux:
%%%%%%%%%%%%%%%%%%%%%%%%%%
\begin{equation}
\Phi^{CC}(\nu_e) = (1.75 \pm 0.15) \times 10^{6}~{\rm cm^{-2}s^{-1}},
\label{phinue}
\end{equation}
%%%%%%%%%%%%%%%%%%%%%%%%%%%%%%%%%%%%%
%
\noindent where we have added the 
statistical and systematic
errors and the estimated theoretical uncertainty
(due to the uncertainty in the CC reaction 
cross section) given in \cite{SNO1} in quadrature.
Utilizing the data on 
$\Phi^{CC}(\nu_e)$ and the
data on the  
solar neutrino flux  
obtained by the Super-Kamiokande
experiment, 
it is possible to  
deduce \cite{SNO1} (see also \cite{FogliSNO}) 
the value of the non-electron neutrino 
component in the flux of solar neutrinos 
measured by the 
Super-Kamiokande collaboration:
%%%%%%%%%%%%%%%%%%%%%%%%%%
\begin{equation}
\Phi(\nu_{\mu,\tau}) = (3.69 \pm 1.13) \times 10^{6}~{\rm cm^{-2}s^{-1}}.
\label{phinumu}
\end{equation}
%%%%%%%%%%%%%%%%%%%%%%%%%%%%%%%%%%%%%
%
\noindent This flux is different from zero at 
more than 3 s. d. 

 Global analyses of the solar neutrino data 
\cite{SKsol,Cl98,Kam96,SAGE,GALLEXGNO},
including the SNO results \cite{SNO1}
and the Super-Kamiokande data on the $e^{-}-$spectrum 
and day-night asymmetry, 
in terms of the neutrino oscillation hypothesis
show   
\cite{FogliSNO,ConchaSNO,GoswaSNO,StrumiaSNO,KSSNO,GiuntiSNO,MSmySNO1,ConchaSNO2,GiuntiSNO2,SKSmySNO} (see also \cite{PeeSNO}) 
that the data favor the large mixing 
angle (LMA) MSW, the LOW and the 
quasi-vacuum oscillation 
(QVO) solutions of the 
solar neutrino problem
with transitions into 
active neutrinos.
In the case of the LMA solution, 
the range of values of the 
neutrino mass-squared difference 
$\Delta m^2 > 0$, characterizing
the two-neutrino transitions of 
the solar neutrinos into an active neutrino, 
$\nu_e \rightarrow \nu_{\mu(\tau)}$,
was found, e.g., 
in \cite{FogliSNO} and \cite{ConchaSNO}
to extend (at 99\% C.L.) to
$\sim 5.0\times 10^{-4}~{\rm eV^2}$
and $\sim 8.0\times 10^{-4}~{\rm eV^2}$, 
respectively: 
%%%%%%%%%%%%%%%%%%%%%%%
\begin{equation}
{\rm LMA~MSW}:~~~~~~2.0\times 10^{-5}~{\rm eV^2} 
\ltap \Delta m^2 
\ltap  
(5.0 - 8.0)\times 10^{-4}~{\rm eV^2}~. 
% ~~ (99\% {\rm C.L.}).
\label{dmsolLMA}
\end{equation}
%%%%%%%%%%%%%%%%%%%%%%%
% 
\noindent The best fit values of $\Delta m^2$  
obtained in the independent analyses 
\cite{FogliSNO,ConchaSNO,GoswaSNO,StrumiaSNO,KSSNO}
are grouped in the narrow interval 
$(\Delta m^2)_{BFV} = (4.3 - 4.9)\times 10^{-5}~{\rm eV^2}$.
A smaller best fit value was found in 
\cite{ConchaSNO2},
$(\Delta m^2)_{BFV} = (3.3 - 3.7)\times 10^{-5}~{\rm eV^2}$,
while a larger value was obtained, e.g., in 
\cite{SKSmySNO}:
$(\Delta m^2)_{BFV} = 6.0\times 10^{-5}~{\rm eV^2}$.
Similar results,
$(\Delta m^2)_{BFV} = 6.3\times 10^{-5}~{\rm eV^2~and~}
6.1\times 10^{-5}~{\rm eV^2}$,
were obtained in \cite{GiuntiSNO} and in \cite{GiuntiSNO2}
by performing a Bayesian analysis 
of the solar neutrino data.
For the mixing parameter
$\sin^22\theta$, which controls the oscillations
of the solar neutrinos, it was found, e.g., 
in \cite{FogliSNO} at 99\% C.L.:
%%%%%%%%%%%%%%%%%%%%%%%
\begin{equation}
{\rm LMA~~MSW}:~~~~~~~~~~~~~~~~~
0.60 \ltap \sin^22\theta \ltap 0.99,~~~~~~~~~~~~~~~~~~~~~~~~~~~~~
\label{thLMA}
\end{equation}
%%%%%%%%%%%%%%%%%%%%%%%%%%%
%
\noindent The best fit values of
$\sin^22\theta$ obtained, e.g., in 
\cite{FogliSNO,ConchaSNO,GoswaSNO,StrumiaSNO,KSSNO}
are confined to the interval   
$(\sin^22\theta)_{BFV} = (0.79 - 0.82)$.
Somewhat smaller values were found 
in \cite{ConchaSNO2}, \cite{GiuntiSNO2} and in \cite{SKSmySNO}:
$(\sin^22\theta)_{BFV} = (0.75 - 0.79);~0.76;~0.77$,
respectively.

  Detailed results were obtained in
\cite{FogliSNO,ConchaSNO,GoswaSNO,StrumiaSNO,KSSNO,GiuntiSNO,ConchaSNO2,GiuntiSNO2,SKSmySNO}
for the LOW solution as well. 
The 95\% C.L. allowed intervals of values 
of $\Delta m^2$ and $\sin^22\theta$ 
found in \cite{FogliSNO}, for instance, 
read:
%%%%%%%%%%%%%%%%%%%%%%%
\begin{equation}
{\rm LOW}:~~~~~~6.0\times 10^{-8}~{\rm eV^2} 
\ltap \Delta m^2
\ltap  
1.8\times 10^{-7}~{\rm eV^2}~,~~~
0.94 \ltap \sin^22\theta \ltap 1.0. 
% ~~ (99\% {\rm C.L.}).
\label{dmsolthLOW}
\end{equation}
%%%%%%%%%%%%%%%%%%%%%%%
%
\noindent The best fit values of
$\Delta m^2$  and $\sin^22\theta$ for the LOW solution,
derived, e.g., 
in \cite{FogliSNO,ConchaSNO,GoswaSNO,StrumiaSNO,KSSNO,ConchaSNO2}
are compatible with each other and are all
approximately given by 
$(\Delta m^2)_{BFV} \cong 10^{-7}~{\rm eV^2}$ and
$(\sin^22\theta)_{BFV} \cong (0.94 - 0.97)$.
A substantially different value
of $(\Delta m^2)_{BFV}$ was
found in \cite{SKSmySNO}:
$(\Delta m^2)_{BFV} \cong 5.5\times 10^{-8}~{\rm eV^2}$ and
$(\sin^22\theta)_{BFV} \cong 0.99$.

   The analyses \cite{FogliSNO,ConchaSNO,GoswaSNO,StrumiaSNO,KSSNO,GiuntiSNO,SKSmySNO}
were based, in particular, on the
standard solar model (SSM) predictions 
of ref. \cite{BPB01} (BP2000) for the 
different components of the solar neutrino flux
($pp$, $pep$, $^7$Be, $^8$B, $CNO$, $hep$, $^{17}$F).
In \cite{FogliSNO,ConchaSNO,GoswaSNO,StrumiaSNO,KSSNO,GiuntiSNO,ConchaSNO2}
the published Super-Kamiokande 
data on the day-night (D - N) asymmetry \cite{SKsol} 
were used as input in the analyses, while 
in \cite{SKSmySNO} the latest (preliminary) results
on the D-N asymmetry, obtained from the
analysis of {\it all currently available
Super-Kamiokande solar neutrino data} 
was utilized (see further). 
% In \cite{GiuntiSNO} the Bayesian method 
% of data analysis was used.
The authors of ref. \cite{ConchaSNO2} have 
used in their analysis 
a new value of the $^8$B neutrino flux
which is suggested by the results of the 
latest (and more precise) 
experimental measurement \cite{pBe701}
of the cross section of the reaction
$p + ^{7}$Be $\rightarrow ^{8}$B $+ \gamma$.
According to the SSM, the $^8$B is produced in the Sun 
in the indicated reaction and 
the $\beta^{+}-$decay of $^8$B in the central part
of the Sun gives rise to the solar $^8$B neutrino flux. 
The results obtained in \cite{pBe701}
give a larger $p - ^7$Be reaction
cross-section (with smaller uncertainty),
and correspondingly - a larger 
astrophysical factor $S_{17}$
(see, e.g., \cite{ConchaSNO2})
than the one used in \cite{BPB01},
which implies, in particular, a larger value of the 
$^8$B neutrino flux than the value predicted 
\footnote{The $^8$B neutrino flux predicted 
in \cite{BPB01} reads 
$\Phi(B)_{BP2000} = 5.05 \times (1 ^{+0.20}_{-0.16})\times 10^{6}~
{\rm cm^{-2}s^{-1}}$, while the flux
utilized in the analysis performed
in \cite{ConchaSNO2} is 
$\Phi(B)_{NEW} = 5.93\times (1 ^{+0.14}_{-0.13})\times 10^{6}~
{\rm cm^{-2}s^{-1}}$.
}
in \cite{BPB01}. In the global Bayesian analysis performed in
\cite{GiuntiSNO2} the SSM predictions for the solar neutrino 
fluxes were not used: both the values of the fluxes and of the
oscillation parameters were derived from the data.

  The best fit values of $\Delta m^2$ 
found in \cite{FogliSNO,ConchaSNO,GoswaSNO,StrumiaSNO,KSSNO}
differ from that derived in \cite{SKSmySNO}
essentially due to the difference in the
Super-Kamiokande data on the D-N asymmetry 
used as input in the corresponding analyses: 
in \cite{SKSmySNO} the latest 
(preliminary) Super-Kamiokande 
result implying a smaller mean value 
of the D-N asymmetry than the  
published one in \cite{SKsol} was utilized.
The smaller possible D-N asymmetry
drives $(\Delta m^2)_{BFV}$
to larger (smaller) value in the 
LMA MSW (LOW) solution region \cite{SKSmySNO}.
Although the data on the D-N asymmetry
used in \cite{FogliSNO,ConchaSNO,GoswaSNO,StrumiaSNO,KSSNO}
and in \cite{ConchaSNO2} are the same,
the best fit value of $\Delta m^2$
in the LMA MSW solution region
found in \cite{ConchaSNO2}
is smaller than those found 
in \cite{FogliSNO,ConchaSNO,GoswaSNO,StrumiaSNO,KSSNO}
because of the difference between the values of the
astrophysical factor $S_{17}$, and
thus of the $^8$B neutrino flux,
used in \cite{ConchaSNO2} and in 
\footnote{Let us note that, e.g., 
in \cite{ConchaSNO,KSSNO,MSmySNO1,ConchaSNO2}
results obtained by treating the $^8$B neutrino flux
as a free parameter in the analysis were also reported.
These results were taken into account when we 
quoted above the $\Delta m^2$ and $\sin^22\theta$
best fit values.
} 
\cite{FogliSNO,ConchaSNO,GoswaSNO,StrumiaSNO,KSSNO}.

    In the present article we update 
our earlier predictions \cite{SK97I,SK97II,DNSNO00}
for the D-N asymmetry
for the Super-Kamiokande and SNO experiments,
taking into account the recent progress 
in the studies of solar neutrinos.
The day-night (D-N) effect -
a difference between the 
solar neutrino event rates
during the day and during the night, 
caused by the additional transitions of the solar
neutrinos taking place at night while the neutrinos 
cross the Earth on the way to the detector 
(see, e.g., \cite{HataL94,DNold} 
and the references quoted therein),
is a unique testable prediction of the 
MSW solutions of the solar neutrino problem. 
The experimental observation of a 
non-zero D-N asymmetry
%%%%%%%%%%%%%%%%%%%%%%%%%%%%%%%%%%%%%%%%%%%%%
\begin{equation}
A^{N}_{D-N} \equiv \frac{R_{N} - R_{D}}{(R_N + R_D)/2},
\end{equation}
%%%%%%%%%%%%%%%%%%%%%%%%%%%%%%%%
%
\noindent where $R_N$ and $R_D$ are, e.g., the  
one year averaged event rates in a given detector 
respectively during the night and the day, 
would be a very strong evidence in favor 
(if not a proof) of an MSW solution
of the solar neutrino problem.
Extensive predictions 
for the magnitude of the D-N effect
for the Super-Kamiokande and SNO 
detectors have been obtained in 
\cite{SK97I,SK97II,DNSNO00,LisiM97,BK97,SK98III,Barger01DN}.
High precision calculations of
% the D-N effect were performed. Extensive predictions for 
the D-N asymmetry in the 
one year averaged recoil-e$^{-}$ spectrum
measured in the Super-Kamiokande experiment
and in the energy-integrated event rates 
for the two experiments were performed
for three event samples, 
{\it Night}, {\it Core} and {\it Mantle},
in \cite{SK97I,SK97II,DNSNO00,SK98III}.
The night fractions of these event samples 
are due to neutrinos which respectively cross 
the Earth along any trajectory, 
cross the Earth core, and
cross only the Earth mantle (but not the core),
on the way to the detector. 

   We focus here, in particular, 
on providing detailed predictions   
for the D-N asymmetry for the LMA MSW and the LOW 
solutions of the solar neutrino problem,
which are favored by the current solar neutrino data.
We will consider in what follows
the {\it Night} (or {\it Full Night})
and the {\it Core} D-N asymmetries,
$A^{N}_{D-N}$ and $A^{C}_{D-N}$. 
The current Super-Kamiokande data \cite{SKsol}
do not contain evidence for 
a substantial D-N asymmetry: the latest published result 
on $A^{N}_{D-N}$ reads \cite{SKsol}
%%%%%%%%%%%%%%%%%%%%%%%%%%%%%%%%%
\begin{equation}
A^{N}_{D-N}(SK) = 0.033 \pm 0.022~(stat.)~^{+0.013}_{-0.012}~(syst.),
\label{dnsk01}
\end{equation}
%%%%%%%%%%%%%%%%%%%%%%%
%
\noindent while the result of the latest analysis
of {\it all currently available Super-Kamiokande 
solar neutrino data} gives even smaller 
mean value \cite{SKSmySNO}
%%%%%%%%%%%%%%%%%%%%%%%%%%%%%%%%%
\begin{equation}
A^{N}_{D-N}(SK) = 0.021 \pm 0.022~(stat.)~^{+0.013}_{-0.012}~(syst.).
\label{dnsk01Smy}
\end{equation}
%%%%%%%%%%%%%%%%%%%%%%%
%
\noindent Adding the errors in eqs.
(\ref{dnsk01}) and (\ref{dnsk01Smy})
in quadrature, one finds that
at 1.5 (2.0) s.d.,
$A^{N}_{D-N}(SK) < 0.072~(0.085)$ 
and $A^{N}_{D-N}(SK) < 0.060~(0.073)$,
respectively.

   We give in the present article 
also detailed predictions for another important 
observable - the ratio
of the event rates of the CC reaction
$\nu_e + D \rightarrow e^{-} + p + p$,
$R_{SNO}(CC)$, and of the neutral current (NC)
reaction $\nu + D \rightarrow \nu + n + p$,   
$R_{SNO}(NC)$, induced by the solar neutrinos in SNO,
%%%%%%%%%%%%%%%%%%%%%%%%%%%%%%%%%%%%%
\begin{equation}
R^{SNO}_{CC/NC} \equiv \frac{\frac{R_{SNO}(CC)}{R_{SNO}(NC)}}
{\frac{R^{0}_{SNO}(CC)}{R^{0}_{SNO}(NC)}}~~,
\label{ccnc}
\end{equation}
%%%%%%%%%%%%%%%%%%%%%%%%%
%
\noindent which is normalized above to 
the value of the same ratio in the absence 
of oscillations of solar neutrinos,
$R^{0}_{SNO}(CC)/R^{0}_{SNO}(NC)$.
First results on the D-N asymmetry 
and on the CC to NC event rate ratio $R^{SNO}_{CC/NC}$
are expected to be published 
in the near future by the SNO collaboration.
We discuss as well the possibilities 
to further constrain the regions of the LMA MSW and
LOW-QVO solutions of the solar neutrino problem
by using the forthcoming SNO data
on the D-N asymmetry  $A^{N}_{D-N}$ and on the 
CC to NC event rate ratio
$R^{SNO}_{CC/NC}$.

  Updated predictions for 
the {\it Night} D-N asymmetry and the 
{\it average} CC to NC event rate ratio
for the SNO experiment were derived after 
the publication of the first SNO results
also in \cite{KSSNO,ConchaSNO2}. 
However, our study overlaps little with 
those performed in \cite{KSSNO,ConchaSNO2}.
%%%%%%%%%%%%%%%%%%%%%%%%%%%%%%%%%%%%%%%
\vspace{-0.5cm}
\section{The LMA MSW and LOW Solutions and the D-N Asymmetry for 
Super-Kamiokande and SNO Experiments}
\vskip -0.2cm
%%%%%%%%%%%%%%%%%%%%%%%%%%%%%%%%%%%%%%%%
\hskip 0.5truecm   Our predictions for the 
{\it Full Night} D-N asymmetry 
in the regions of the LMA MSW and LOW solutions
of the solar neutrino problem 
for the Super-Kamiokande and SNO experiments,
$A^{N}_{D-N}(SK)$ and $A^{N}_{D-N}(SNO)$,
are shown in Figs. 1 (upper panel) and 2, respectively,
while in Figs. 1 (lower panel) and 3 we show predictions for the 
{\it Core} D-N asymmetry for the two detectors
\footnote{The calculations 
of the D-N effect for the Super-Kamiokande and 
SNO detectors performed in the 
present article are based on 
the methods developed for our
earlier studies of the D-N effect for these
detectors, which are described 
in detail in \cite{SK97I,SK97II,DNSNO00}.
Here we use the BP2000 SSM \cite{BPB01}
predictions for the electron number density
distribution in the Sun.
},
$A^{C}_{D-N}(SK)$ and $A^{C}_{D-N}(SNO)$.
The calculations of $A^{N,C}_{D-N}(SNO)$ 
have been performed 
by taking into account, in particular, 
the energy resolution function
of the SNO detector \cite{SNO1}. 
The effect of the energy resolution function
on the values of the {\it Full Night} 
and {\it Core} D-N asymmetries, 
$A^{N,C}_{D-N}(SNO)$, as our results show, 
is negligible for values of the asymmetries   
$A^{N,C}_{D-N}(SNO) \geq 0.01$. 
In Figs. 2 and 3 we show contours of
constant $A^{N}_{D-N}(SNO)$ and
$A^{C}_{D-N}(SNO)$ in the 
% $\Delta m^2 - \sin^22\theta$
$\Delta m^2 - \tan^2\theta$
plane for two values of the
threshold kinetic energy of the 
final state electron
\footnote{The results of our calculations show
that in the LMA MSW and LOW solution regions
the predicted {\it Mantle} D-N asymmetry
in the CC event rate at SNO
\cite{DNSNO00} practically
coincides with the
{\it Full Night} D-N asymmetry.}, 
$T_{\mathrm{e},\mathrm{th}} = 6.75$\ MeV
(upper panels) and 5.0 MeV (lower panels).
The published SNO data were obtained 
using the first value \cite{SNO1}, 
while the second one is the 
threshold energy planned 
to be reached at a later stage
of the experiment. A comparison of the upper 
and lower panels in Fig. 2 shows that 
the {\it Full Night} D-N asymmetry
$A^{N}_{D-N}(SNO)$ {\it decreases} somewhat
in the LMA MSW solution region -
approximately by $\sim (8 - 10)\%$, when
$T_{\mathrm{e},\mathrm{th}}$  
is decreased from 6.75 MeV to 5.0 MeV.
The change in the LOW solution region
is opposite and 
larger in magnitude than in the LMA MSW solution region:
$A^{N}_{D-N}(SNO)$ {\it increases} by about
$\sim (15 - 20)\%$ when
$T_{\mathrm{e},\mathrm{th}}$ is reduced
from 6.75 MeV to 5.0 MeV.
The above results imply that
for given $\sin^22\theta$,
the same values of the asymmetry 
at $T_{\mathrm{e},\mathrm{th}} = 5.0$\ MeV 
occur in both the LMA MSW and LOW solution regions
at smaller values of $\Delta m^2$
than for $T_{\mathrm{e},\mathrm{th}} = 6.75$\ MeV.   
Qualitatively similar conclusions are valid for the 
{\it Core} D-N asymmetry 
$A^{C}_{D-N}(SNO)$ (Fig. 3).
  
   Consider the predictions for the D-N 
asymmetry in the case of the LMA MSW solution.
As Figs. 1 - 3 show, at 
$\Delta m^2 \gtap 1.5\times 10^{-4}~{\rm eV^2}$
both $A^{N}_{D-N}(SK)$ and $A^{N,C}_{D-N}(SNO)$
are smaller than 1\%. For given 
$\Delta m^2 \ltap 10^{-4}~{\rm eV^2}$
and $\sin^22\theta$ 
from the LMA solution region we have
\cite{DNSNO00} 
$A^{N}_{D-N}(SNO) \cong (1.5 - 2.0)A^{N}_{D-N}(SK)$.
The difference between 
$A^{N}_{D-N}(SK)$ and $A^{N}_{D-N}(SNO)$
in the indicated region is due 
to i) the contribution of the NC
$\nu_{\mu(\tau)} - e^{-}$ 
elastic scattering reaction
(in addition to that due to the
$\nu_{e} - e^{-}$ elastic scattering) 
to the solar neutrino event rate
measured by the Super-Kamiokande
experiment,
and ii) to the relatively small value 
of the solar $\nu_e$ survival 
probability in the Sun, $\bar{P} \sim 0.3$.
The indicated NC contribution to the
Super-Kamiokande event rate
tends to diminish the D-N asymmetry.
Obviously, there is no similar contribution
to the SNO CC event rate.

  Thus, in the case of the LMA MSW 
solution of the solar neutrino
problem, the D-N asymmetry measured in SNO
can be considerably larger than the D-N asymmetry 
measured in the Super-Kamiokande
experiment \cite{DNSNO00}. 
The 2 s.d. upper limit on
the D-N asymmetry, 
$A^{N}_{D-N}(SK) < 8.5\%~(7.3\%)$,
following from the Super-Kamiokande data,
eq. (7) (eq. (8)), for instance, does not exclude
a D-N asymmetry in the SNO CC event rate
as large as $\sim (10 - 15)\%$.
As Fig. 2 shows, $A^{N}_{D-N}(SNO)$ 
can reach a value of $\sim 20\%$
in the 99\% C.L. region of the
LMA MSW solution, eq. (3). 
In the 95\% C.L. LMA solution region
of \cite{SKSmySNO} one has
$A^{N}_{D-N}(SK) \ltap 13\%$.
In the best fit point 
of the LMA MSW solution, found in 
\cite{FogliSNO,ConchaSNO,GoswaSNO,KSSNO},
we get for $T_{\mathrm{e},\mathrm{th}} = 
6.75~{\rm MeV~(~5.0~MeV})$,
$(A^{N}_{D-N}(SNO))^{LMA}_{BF1} \cong 7.3~(6.6)\%$.
% 7.1~(6.6)\%.
Even larger value of the asymmetry 
$A^{N}_{D-N}(SNO)$ corresponds to the 
best fit point obtained in 
\cite{ConchaSNO2}: 
$(A^{N}_{D-N}(SNO))^{LMA}_{BF2} \cong 10.1~(9.3)\%$.
At the same time, one finds a considerably 
smaller value of $A^{N}_{D-N}(SNO)$ in the 
best fit point derived in \cite{SKSmySNO}:
$(A^{N}_{D-N}(SNO))^{LMA}_{BF3} \cong 5.0~(4.6)\%$.
The {\it Full Night} D-N asymmetry
in the Super-Kamiokande detector
in the indicated three different 
best fit points found in
\cite{FogliSNO,ConchaSNO,GoswaSNO,KSSNO},
\cite{ConchaSNO2} and \cite{SKSmySNO} for
$T_{\mathrm{e},\mathrm{th}} = 5.0~{\rm MeV}$
read, respectively: 
 $(A^{N}_{D-N}(SK))^{LMA}_{BF} \cong 3.9\%;~5.4\%;~2.6\%$.
% 3.6\%;~5.0\%;~2.5\%. 

  Actually, as it is not difficult to show,
the following approximate relation between
$A^{N}_{D-N}(SK)$ and $A^{N}_{D-N}(SNO)$
holds for fixed 
$\Delta m^2$ and $\sin^22\theta$ 
from the region of the LMA MSW solution:
%%%%%%%%%%%%%%%%%%%%%%%%%%
\begin{equation}
A^{N}_{D-N}(SNO) \cong A^{N}_{D-N}(SK)
\left [ 1 + \frac{r}{(1-r)\bar{P}} \right ]~,
\label{SKSNODN} 
\end{equation}
%%%%%%%%%%%%%%%%%%%%%%%%%%%%%%%%
%
 \noindent where $r \equiv 
\sigma(\nu_{\mu(\tau)}e^{-})/\sigma(\nu_{e}e^{-})$,
$\sigma(\nu_{l}e^{-})$
being the $\nu_{l} - e^{-}$ elastic 
scattering cross section, $l=e,\mu,\tau$,
and $\bar{P}$ is the
average probability of solar
$\nu_e$ survival in the Sun.
For the solar neutrino energies of 
interest one has $r \cong 0.155$.
For $\Delta m^2$ and $\sin^22\theta$ 
from the LMA MSW solution region,
the transitions of the solar ($^8$B) neutrinos
with energies $E \gtap 5.0$ MeV are
adiabatic and in a relatively 
large sub-region one finds
$\bar{P} \cong \sin^2\theta$. We would like 
to emphasize that the relation (\ref{SKSNODN})
is not very precise and can serve only for
rough estimates.

  As a comparison of Figs. 2 and 3 indicates,
for given $\Delta m^2$ and $\sin^22\theta$
from the LMA MSW solution region,
the {\it Core} D-N asymmetry 
in the SNO detector is 
predicted to be larger
than the {\it Full Night}
D-N asymmetry typically by a 
factor of $\sim 1.2$ \cite{DNSNO00}:
$A^{C}_{D-N}(SNO) \cong 1.2A^{N}_{D-N}(SNO)$.

 The predicted values of 
$A^{N}_{D-N}(SK)$ and $A^{N}_{D-N}(SNO)$
in the LOW solution region differ 
less than in the case of the
LMA MSW solution since the 
average survival probability
$\bar{P}$ is typically by a factor
of $\sim 1.5$ larger  
for the LOW solution than in the case of
the LMA MSW solution.
As it follows from Fig. 2,
in the LOW solution region given
by eq. (5) one has
$(A^{N}_{D-N}(SNO))^{LOW} \cong (1.0 - 7.5)\%$.
A comparison of Figs. 2 and 3 
allows to conclude that
in the region under discussion we have
$(A^{C}_{D-N}(SNO))^{LOW} \cong (A^{N}_{D-N}(SNO))^{LOW}$.
In the best fit point of the solution's
region found in \cite{FogliSNO,ConchaSNO,GoswaSNO,KSSNO}
and in \cite{ConchaSNO2} 
we get for $T_{\mathrm{e},\mathrm{th}} = 
6.75~{\rm MeV~(~5.0~MeV})$,
$(A^{N}_{D-N}(SNO))^{LOW}_{BF1,2} \cong 3.8~(4.2)\%$,
% 3\%,
while in the best fit point 
obtained in \cite{SKSmySNO} one has
$(A^{N,C}_{D-N}(SNO))^{LOW}_{BF3} \cong 1.2~(1.5)\%$.
% < 0.5\%. 
Similar predictions are valid for
$A^{N}_{D-N}(SK)$ (Fig. 1). 
Obviously, an observation
of $A^{N}_{D-N}(SNO) \gtap 10\%$
will strongly disfavor 
the LOW solution of the solar neutrino problem.

  As Fig. 2 indicates,
an observation of a non-zero D-N asymmetry
which is definitely greater than 1\%,
$A^{N}_{D-N}(SNO) > 1\%$, would rule out the
QVO solution
which requires values of $\Delta m^2$
from the interval 
$\Delta m^2 \sim (5\times 10^{-10} - 5\times 10^{-8})~{\rm eV^2}$  
and $\sin^22\theta \cong (0.70 - 1.0)$
(for a discussion of the QVO oscillations 
of solar neutrinos and of the QVO
solution see, e.g., 
\cite{FogliSNO,ConchaSNO,GoswaSNO,StrumiaSNO,KSSNO,GiuntiSNO,LMMPPQVO}).
%%%%%%%%%%%%%%%%%%%%%%%%%%%%%%%%%%%%%%%%%%%%%%%%%%%%%
\vspace{-0.5cm}
\section{Predictions for $R^{SNO}_{CC/NC}$}
\vskip -0.2cm
%%%%%%%%%%%%%%%%%%%%%%%%%%%%%%%%%%%%%%%%%%%%%%%%
\hskip 0.5cm The importance of the measurement of
the CC to NC solar neutrino event rate ratio
in the SNO experiment,
$R^{SNO}_{CC/NC}$, for determining 
the correct solution of the solar neutrino problem 
has been widely discussed 
(see, e.g., 
\cite{CCNCSNO,BKSCCNC01} 
and the references quoted therein).
We have performed a high precision
calculations of the ratio 
$R^{SNO}_{CC/NC}$, in particular,
for three different
CC reaction cross-sections
which were taken from
\cite{XS:IAUTH0,XS:IAUTH1,XS:IAUTH2},
and using the electron number density
distribution in the Sun from \cite{BPB01}.
The differences in the results obtained
for $R^{SNO}_{CC/NC}$ 
using the three cross sections are
negligible in the regions of the
LMA MSW and LOW solutions 
of the solar neutrino problem
of interest. We have found that 
the effect of the
SNO energy resolution 
function on the predictions
for $R^{SNO}_{CC/NC}$ 
is negligible as well.

  The SNO experiment will measure 
the CC and NC average event rates, 
$R_{SNO}^{exp}(CC)$ and $R_{SNO}^{exp}(NC)$.
In order to compare these results with 
the predictions for the double ratio
$R^{SNO}_{CC/NC}$, eq. (\ref{ccnc}),
one has to use as a normalization factor
the theoretically calculated (in the absence
of solar neutrino oscillations)
value of the ratio $R^{0}_{SNO}(CC)/R^{0}_{SNO}(NC)$.
The ratio $R^{0}_{SNO}(CC)/R^{0}_{SNO}(NC)$
of interest is practically the same
when it is calculated within
a given theoretical model,
ref. \cite{XS:IAUTH1} or ref. \cite{XS:IAUTH2},
for the CC and NC reaction cross sections:
for $T_{\mathrm{e},\mathrm{th}} = 5.00~(6.75)$ MeV we
find $R^{01}_{SNO}(CC)/R^{01}_{SNO}(NC) = 1.927~(1.232)$
using the CC and NC cross sections derived
in ref. \cite{XS:IAUTH1}; utilizing the results of
ref. \cite{XS:IAUTH2} we get:
$R^{02}_{SNO}(CC)/R^{02}_{SNO}(NC) = 1.933~(1.235)$.
This is not the case, however, if one calculates
the ratio of interest
% $R^{0}_{SNO}(CC)/R^{0}_{SNO}(NC)$
by taking the CC reaction cross section from 
ref. \cite{XS:IAUTH1}
and the NC reaction cross section from
ref. \cite{XS:IAUTH2} and vice versa -
the CC cross section from 
ref. \cite{XS:IAUTH2} and the NC cross section
from ref. \cite{XS:IAUTH1}.
One finds for $T_{\mathrm{e},\mathrm{th}} = 5.00~(6.75)$ MeV
in the two cases, respectively:
$R^{01}_{SNO}(CC)/R^{02}_{SNO}(NC) = 2.049~(1.310)$
and $R^{02}_{SNO}(CC)/R^{01}_{SNO}(NC) = 1.818~(1.161)$. 
Now the relative difference between 
the two calculated ratios is $\sim (10 - 15)\%$ and
cannot be neglected. This suggests, in particular,
that the ratio 
$R^{0}_{SNO}(CC)/R^{0}_{SNO}(NC)$
should be calculated within
a given theoretical model for the
CC and NC reaction cross sections. 
In what regards the most recent calculations of the 
CC and NC reaction cross sections \cite{Nakamura01},
they lead to a value of the ratio
$R^{0}_{SNO}(CC)/R^{0}_{SNO}(NC)$ 
which does not differ from that obtained in
ref. \cite{XS:IAUTH1} by more than $1\%$.

   Similarly, although 
the effect of the SNO energy resolution 
function on the predictions
for the double ratio $R^{SNO}_{CC/NC}$ 
is negligible, it is not negligible
in the case of the ratio
$R^{0}_{SNO}(CC)/R^{0}_{SNO}(NC)$.
For $T_{\mathrm{e},\mathrm{th}} = 6.75$ MeV,
for instance, we find that the value of
$R^{0}_{SNO}(CC)/R^{0}_{SNO}(NC)$
calculated using the results of 
ref. \cite{XS:IAUTH1}
(or of ref. \cite{XS:IAUTH2})
assuming ideal resolution, 
is by a factor of 1.033 bigger than
the value obtained by taking the
SNO resolution function into account. 

  Our predictions for the average ratio 
during the day ({\it Day} ratio), $R^{SNO}_{CC/NC}(D)$,
during the night ({\it Full Night} ratio), $R^{SNO}_{CC/NC}(N)$,
and for the case of the
CC event rate produced at night 
by solar neutrinos which cross the Earth core
on the way to SNO ({\it Core} ratio), 
$R^{SNO}_{CC/NC}(C)$,
are shown respectively in Figs. 4 - 6.
Results for each of the three ratios were
obtained for two values of the 
(effective) kinetic energy 
threshold of the detected $e^{-}$ 
in the CC reaction:
for $T_{\mathrm{e},\mathrm{th}} = 6.75$\ MeV 
(upper panels) and
$T_{\mathrm{e},\mathrm{th}} = 5.00$\ MeV (lower
panels). 

  A comparison of Fig. 4 and Figs. 5 - 6 
shows that CC to NC ratio 
increases substantially during the night 
for values of $\Delta m^2$ and
$\sin^22\theta$ from the region 
$2\times 10^{-7}~{\rm eV^2}\ltap \Delta m^2 
\ltap  2\times 10^{-5}~{\rm eV^2}$,
$10^{-2} \ltap \sin^22\theta \ltap 0.98$,
which, however, is not favored by the 
current solar neutrino data.
The increase is due to the Earth matter effect.
The difference between {\it Night} and
{\it Core} ratio in the indicated region
is essentially caused by the Earth mantle-core 
interference effect \cite{SP98}.
For $\Delta m^2 >  2\times 10^{-5}~{\rm eV^2}$
in the LMA solution region, 
and in all the LOW solution region,
the difference between the {\it Core} and
and {\it Night} ratios is negligible,
$R^{SNO}_{CC/NC}(N) \cong R^{SNO}_{CC/NC}(C)$.
At $\Delta m^2 \gtap 8\times 10^{-5}~{\rm eV^2}$
in the LMA region, 
and at $\Delta m^2 \ltap 10^{-7}~{\rm eV^2}$ 
in the LOW-QVO region,
the {\it Day} and the {\it Night} ratios 
practically coincide,
$R^{SNO}_{CC/NC}(D) \cong R^{SNO}_{CC/NC}(N)$.

   As the results exhibited in Figs. 4 - 6 indicate,
when $T_{\mathrm{e},\mathrm{th}}$ is decreased
from 6.75  MeV to 5.0 MeV,
the three ratios $R^{SNO}_{CC/NC}(X)$, $X=D,N,C$, 
change little and only 
in relatively small 
sub-regions 
of the LMA MSW and of the LOW
solution regions
(compare the upper and lower
panels in each of Figs. 4 - 6).
In the best fit points
of the LMA MSW and LOW 
solutions, the three ratios 
$R^{SNO}_{CC/NC}(X)$, $X=D,N,C$,
do not change when the threshold
energy is reduced from 6.75  MeV to 5.0 MeV
(see further).

 In the 99\% C.L. LMA MSW solution region,
eqs. (\ref{dmsolLMA}) and (\ref{thLMA}), 
we find that each of the {\it Day},
{\it Night} and {\it Core} ratios 
$R^{SNO}_{CC/NC}(X)$, $X=D,N,C$,
can take values in the interval
$R^{SNO}_{CC/NC}(X) \cong (0.20 - 0.65)$
for both $T_{\mathrm{e},\mathrm{th}} = 6.75~{\rm MeV}$
and $T_{\mathrm{e},\mathrm{th}} = 5.00~{\rm MeV}$.
If $\Delta m^2 \ltap 2\times 10^{-4}~{\rm eV^2}$,
which corresponds to the 95\% C.L. 
solution's region of ref. \cite{FogliSNO}, we have
$R^{SNO}_{CC/NC}(X) \cong (0.20 - 0.45)$,
$X=D,N,C$. In the best fit points 
in the LMA MSW solution region, 
obtained in \cite{FogliSNO,ConchaSNO,GoswaSNO,KSSNO},
\cite{ConchaSNO2} and \cite{SKSmySNO}, 
we get, respectively,
$R^{SNO}_{CC/NC}(D) \cong 0.29;~0.28;~0.27$,
$R^{SNO}_{CC/NC}(N) \cong 0.31;~0.29;~0.29$,
$R^{SNO}_{CC/NC}(C) \cong 0.31;~0.30;~0.30$,
for both values of $T_{\mathrm{e},\mathrm{th}}$,
$T_{\mathrm{e},\mathrm{th}} = 6.75~{\rm MeV};~5.00~{\rm MeV}$.
The ``best fit'' ratios are very sensitive to the
best fit value of $\sin^22\theta$.

  In the case of the LOW solution,
the interval of possible values of
$R^{SNO}_{CC/NC}(X)$, $X=D,N,C$, 
is much narrower if 
$\Delta m^2$ and $\sin^22\theta$ 
lie within the region given by eq. (\ref{dmsolthLOW}):
$R^{SNO}_{CC/NC}(X) \cong (0.38 - 0.45)$.
Somewhat larger values of $R^{SNO}_{CC/NC}(X)$ -
up to $\sim 0.55$,
are possible in the 99.73\% C.L. LOW solution regions
derived in refs. \cite{ConchaSNO,SKSmySNO}.
In the LOW solution best fit points 
found in \cite{FogliSNO,ConchaSNO,GoswaSNO,KSSNO,ConchaSNO2}
and \cite{SKSmySNO}, we obtain for 
$T_{\mathrm{e},\mathrm{th}} = 6.75~{\rm MeV}~(5.00~{\rm MeV})$, 
respectively, $R^{SNO}_{CC/NC}(D) \cong 0.44~(0.43);~0.49$,
$R^{SNO}_{CC/NC}(N,C) \cong 0.45~(0.44);~0.49$.

  If the average probability of survival
of the solar ($^8$B) $\nu_e$ with energy 
$8.2~{\rm MeV} \ltap E$
$ \ltap 14.0$ MeV
(the flux of which was measured 
by the SNO experiment \cite{SNO1}) 
does not exhibit i) a strong 
dependence on the neutrino energy and ii) a large
day-night variation,
we have, as it is not difficult to show,
%%%%%%%%%%%%%%%%%%%%%%%%%%%%%%%%%%%%%
\begin{equation}
R^{SNO}_{CC/NC} \cong 
\frac{\Phi^{CC}(\nu_e)}{\Phi^{CC}(\nu_e) + 
\Phi(\nu_{\mu,\tau})} \cong 0.32 \pm 0.07,
\label{ccncSNO}
\end{equation}
%%%%%%%%%%%%%%%%%%%%%%%%%
%
\noindent where $R^{SNO}_{CC/NC}$
is the averaged ratio over the period 
of SNO data-taking \cite{SNO1}, and 
we have used eqs. (\ref{phinue}) and
(\ref{phinumu}). Taking into account 
an uncertainty corresponding to 
`` 1 standard deviation'' and
to ``2 standard deviations'',
we find from eq. (\ref{ccncSNO}), respectively,
$0.25 \leq R^{SNO}_{CC/NC} \leq 0.39$
and $0.18 \leq R^{SNO}_{CC/NC} \leq 0.46$.
As Figs. 4 - 5 indicate,
an upper limit $R^{SNO}_{CC/NC} \leq 0.45$
would imply that in the cases of the
LMA MSW and of the LOW solutions one has
$\Delta m^2 \ltap 2\times 10^{-4}~{\rm eV^2}$
and
$\Delta m^2 \gtap 6\times 10^{-8}~{\rm eV^2}$,
respectively.
%%%%%%%%%%%%%%%%%%%%%%%%%%%%%%%%%%%
\vspace{-0.5cm}
\section{Constraining the Solar Neutrino Oscillation Parameters}
\vskip -0.2cm
%%%%%%%%%%%%%%%%%%%%%%%%%%%%%%%%%%%
\hskip 0.5cm It should be clear from 
the discussions in Sections 2 and 3 
that a measured value of 
$A^{N}_{D-N}(SNO) > 1.0\%$
and/or of $R^{SNO}_{CC/NC} \ltap 0.45$
in the SNO experiment can strongly 
diminish the regions of the allowed values of
% $\Delta m^2$ and $\sin^22\theta$
$\Delta m^2$ and $\tan^2\theta$
of the LMA MSW and of 
the LOW solutions of the
solar neutrino problem.
As it follows from the results 
shown graphically in Fig. 2
and in Figs. 4 - 5, 
an experimental upper limit on
$A^{N}_{D-N}(SNO)$
in the case of the LMA MSW 
(LOW) solution
would imply a lower (upper)
limit on $\Delta m^2$.
At the same time, an
experimental upper limit on
$R^{SNO}_{CC/NC}$ would lead to 
an upper (lower) limit on 
$\Delta m^2$. Thus, even upper limits
on $A^{N}_{D-N}(SNO)$ of the order
of 10\% and on $R^{SNO}_{CC/NC}$
of the order of 
% 0.50
0.45 can significantly reduce
the LMA MSW and the LOW
solution regions. 
%%%%%%%%%%%%%%%%%%%%%%%%%%%%%%%%%%%%
\vspace{-0.5cm}
\section{Conclusions}
\vskip -0.2cm
%%%%%%%%%%%%%%%%%%%%%%%%%%%%%%%%%%%%
\hskip 0.5cm  In the present article we have 
derived detailed predictions   
for the D-N asymmetry 
in the solar neutrino induced
CC event rate in the SNO detector
for the LMA MSW and the LOW 
solutions of the solar neutrino problem,
which are favored by the current 
solar neutrino data.
We have obtained results for 
the {\it Night} (or {\it Full Night})
and the {\it Core} D-N asymmetries for SNO,
$A^{N}_{D-N}(SNO)$ and $A^{C}_{D-N}(SNO)$,
which are presented in the form of
iso-(D-N) asymmetry contour plots in the 
% $\Delta m^2 - \sin^22\theta$ plane
$\Delta m^2 - \tan^2\theta$ plane
in Figs. 2 - 3.
Detailed predictions for the {\it Night}
and {\it Core}
D-N asymmetries for the Super-Kamiokande detector,
$A^{N,C}_{D-N}(SK)$, were also derived (Fig. 1).
The high precision calculations of $A^{N,C}_{D-N}(SNO)$ 
have been performed 
by taking into account, in particular, 
the energy resolution function
of the SNO detector \cite{SNO1}. 
Our results show, however, that
the effect of the energy resolution function
on the predicted values of the {\it Full Night} 
and {\it Core} D-N asymmetries
is negligible when
$A^{N,C}_{D-N}(SNO)\geq 0.01$.
The asymmetries $A^{N,C}_{D-N}(SNO)$ 
are calculated for two values of the
threshold (effective) kinetic energy of the 
final state electron,
$T_{\mathrm{e},\mathrm{th}} = 6.75$\ MeV
and 5.0 MeV. The published SNO data were obtained 
using the first value \cite{SNO1}, 
while the second one is the threshold energy planned 
to be reached at a later stage
of the experiment.

  The {\it Full Night} D-N asymmetry in the 
CC event rate in the SNO detector,
$A^{N}_{D-N}(SNO)$, can be 
in the LMA MSW solution region
by a factor of $\sim (1.5 - 2.0)$ 
bigger than the 
{\it Full Night} D-N asymmetry
in the solar neutrino induced
event rate in the Super-Kamiokande 
detector \cite{DNSNO00}: 
$(A^{N}_{D-N}(SNO))^{LMA} \cong 
(1.5 - 2.0) (A^{N}_{D-N}(SK))^{LMA}$.
The asymmetry $A^{N}_{D-N}(SNO)$
measured in the SNO experiment
can be as large as $(15 - 20)\%$.
A value of $A^{N}_{D-N}(SNO) \cong 15\%$,
for instance, cannot be excluded 
by the 95\% C.L. (2 s.d.) upper limit
on $A^{N}_{D-N}(SK)$ 
following from the Super-Kamiokande data
on the D-N effect \cite{SKsol,SKSmySNO}.
In the best fit point of the LMA MSW
solution region found in 
\cite{FogliSNO,GoswaSNO,ConchaSNO,KSSNO}
and in \cite{ConchaSNO2}
we get for $T_{\mathrm{e},\mathrm{th}} = 
6.75~{\rm MeV~(~5.0~MeV})$,
$(A^{N}_{D-N}(SNO))^{LMA}_{BF1} \cong 7.3~(6.6)\%$
and $(A^{N}_{D-N}(SNO))^{LMA}_{BF2} \cong 10.1~(9.3)\%$,
respectively.
At the same time, one finds a considerably 
smaller value of $A^{N}_{D-N}(SNO)$ in the 
LMA solution best fit point obtained in \cite{SKSmySNO}:
$(A^{N}_{D-N}(SNO))^{LMA}_{BF3} \cong 5.0~(4.6)\%$.
In the LMA MSW solution region,
the {\it Core} D-N asymmetry 
in the SNO detector is 
predicted to be larger
than the {\it Full Night}
D-N asymmetry typically by a 
factor of $\sim 1.2$:
$(A^{C}_{D-N}(SNO))^{LMA} \cong 1.2(A^{N}_{D-N}(SNO))^{LMA}$.

 In the case of the 
LOW solution of the solar
neutrino problem one has (Figs. 1 and 2)
in the region where 
$A^{N}_{D-N}(SK) > 1\%$:
$A^{N}_{D-N}(SNO) \cong 
(1.2 - 1.4)A^{N}_{D-N}(SK)$.
In the  solution region given
by eq. (5) we find
$(A^{N}_{D-N}(SNO))^{LOW} \cong (1.0 - 7.5)\%$.
In the region under discussion,
$(A^{C}_{D-N}(SNO))^{LOW} \cong (A^{N}_{D-N}(SNO))^{LOW}$.
In the best fit point of the LOW solution
found in \cite{FogliSNO,ConchaSNO,GoswaSNO,KSSNO}
and in \cite{ConchaSNO2} 
we get $(A^{N}_{D-N}(SNO))^{LOW}_{BF1,2} \cong 3.8~(4.2)\%$
for $T_{\mathrm{e},\mathrm{th}} = 
6.75~{\rm MeV~(~5.0~MeV})$,
while in the best fit point 
obtained in \cite{SKSmySNO} one has
$(A^{N,C}_{D-N}(SNO))^{LOW}_{BF3} \cong 1.2~(1.5)\%$.
% < 0.5\%. 
An observation
of $A^{N}_{D-N}(SNO) \gtap 10\%$
will strongly disfavor 
the LOW solution of the solar neutrino problem, while
an observation of
% a non-zero D-N asymmetry
% which is definitely greater than 1\%,
$A^{N}_{D-N}(SNO) > 1\%$ would rule out the
QVO solution.

  We have derived also 
detailed predictions for the ratio
of the event rates of the CC reaction
$\nu_e + D \rightarrow e^{-} + p + p$,
$R_{SNO}(CC)$, and of the neutral current (NC)
reaction $\nu + D \rightarrow \nu + n + p$,   
induced by the solar neutrinos in SNO
during the {\it day}, $R^{SNO}_{CC/NC}(D)$,
during the {\it night}, $R^{SNO}_{CC/NC}(N)$,
and for the case of the
CC event rate produced at night 
by solar neutrinos which cross the Earth {\it core}, 
$R^{SNO}_{CC/NC}(C)$ (Figs. 4 - 6).
The predictions were 
obtained for $T_{\mathrm{e},\mathrm{th}} = 
6.75~{\rm MeV~and~5.0~MeV}$.
We find that in the LMA MSW solution region
given by eqs. (\ref{dmsolLMA}) and (\ref{thLMA}),
$R^{SNO}_{CC/NC}(X) \cong (0.20 - 0.65)$,
$X=D,N,C$; for $\Delta m^2 \ltap 2\times 10^{-4}~{\rm eV^2}$
from this region
we have $R^{SNO}_{CC/NC}(X) \cong (0.20 - 0.45)$.
In the LOW solution region 
given by eq. (\ref{dmsolthLOW}) we obtain
$R^{SNO}_{CC/NC}(X) \cong (0.38 - 0.45)$.
% ; in the region found in \cite{SKSmySNO}
% the three ratios can take values up to .
In the LMA solution best fit points (see the text)
we get $R^{SNO}_{CC/NC}(X) \cong (0.27 - 0.31)$,
while in the two LOW solution best fit points
discussed in the text we find approximately 
$R^{SNO}_{CC/NC}(X) \cong 0.44~{\rm and}~0.49$.

    In the case of the LMA and LOW solutions,
the value of $A^{N}_{D-N}(SNO)$ is very sensitive
to the value of $\Delta m^2$, while
$R^{SNO}_{CC/NC}$ exhibits a very strong 
dependence on $\tan^2\theta$.
A measured value of 
$A^{N}_{D-N}(SNO) > 1.0\%$
and/or of $R^{SNO}_{CC/NC} \ltap 0.45$
% $\leq 0.50$
in the SNO experiment can strongly 
diminish the regions of the allowed values of
% $\Delta m^2$ and $\tan^2\theta$
$\Delta m^2$ and $\tan^2\theta$
of the LMA MSW and of 
the LOW-QVO solutions of the
solar neutrino problem.
An upper limit on
$A^{N}_{D-N}(SNO)$
in the case of the LMA MSW 
(LOW-QVO) solution
would imply a lower (upper)
limit on $\Delta m^2$.
At the same time, an
experimental upper limit on
$R^{SNO}_{CC/NC}$ would lead to 
an upper (lower) limit on 
$\Delta m^2$. Thus, even upper limits
on $A^{N}_{D-N}(SNO)$ of the order
of 10\% and on $R^{SNO}_{CC/NC}$
of the order of 
% 0.50
0.45 can significantly reduce
the LMA MSW and the LOW-QVO
solution regions. 
%%%%%%%%%%%%%%%%%%%%%%%%%%%%%%%%%%%%%%%
\vspace{-0.5cm} 
\section{Acknowledgements}
\vskip -0.2cm
%%%%%%%%%%%%%%%%%%%%%%%%%%%%%%%%%%%%%%%%
\hskip 0.5cm S.T.P. would like to thank the organizers  
of the International Workshop on Neutrino Oscillations
in Venice (July 24 - 26, 2001), where some of the 
results of the present study were first reported, 
and especially Prof. M. Baldo Ceolin,
for providing excellent conditions for a 
timely and fruitful Workshop. The work of S.T.P. was 
supported in part by the EEC grant HPRN-CT-2000-00152.

\newpage

%%%%%%%%%%%%%%%%%%%%%%%%%%%%
\begin{figure}[ht]
  \centering
\vspace*{13pt}
%  \leftline{\hfill\vbox{\hrule width 5cm height0.001pt}\hfill}
         \mbox{\epsfig{figure=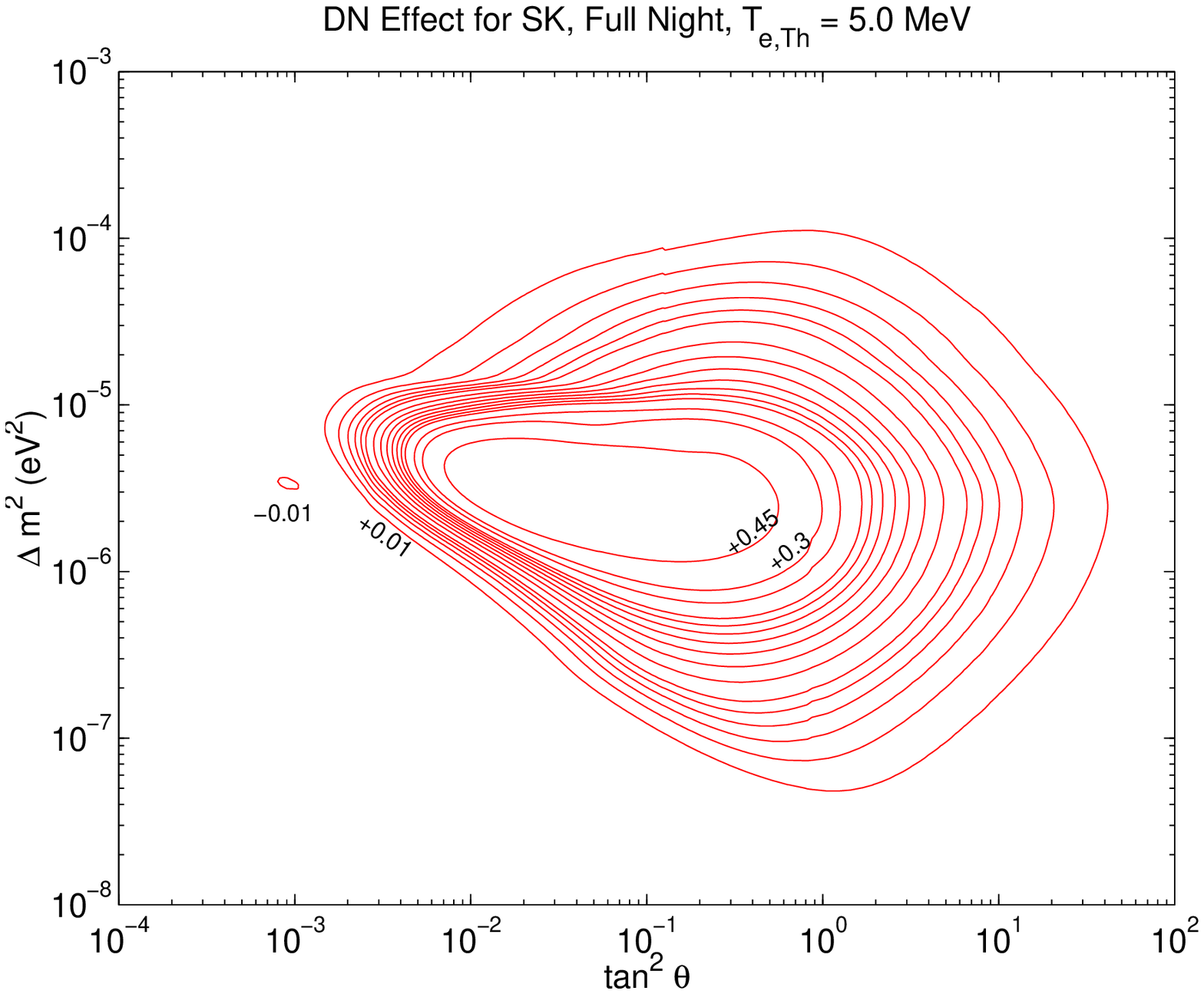,width=12.5cm}}\\
%  \leftline{\hfill\vbox{\hrule width 5cm height0.001pt}\hfill}
       \mbox{\epsfig{figure=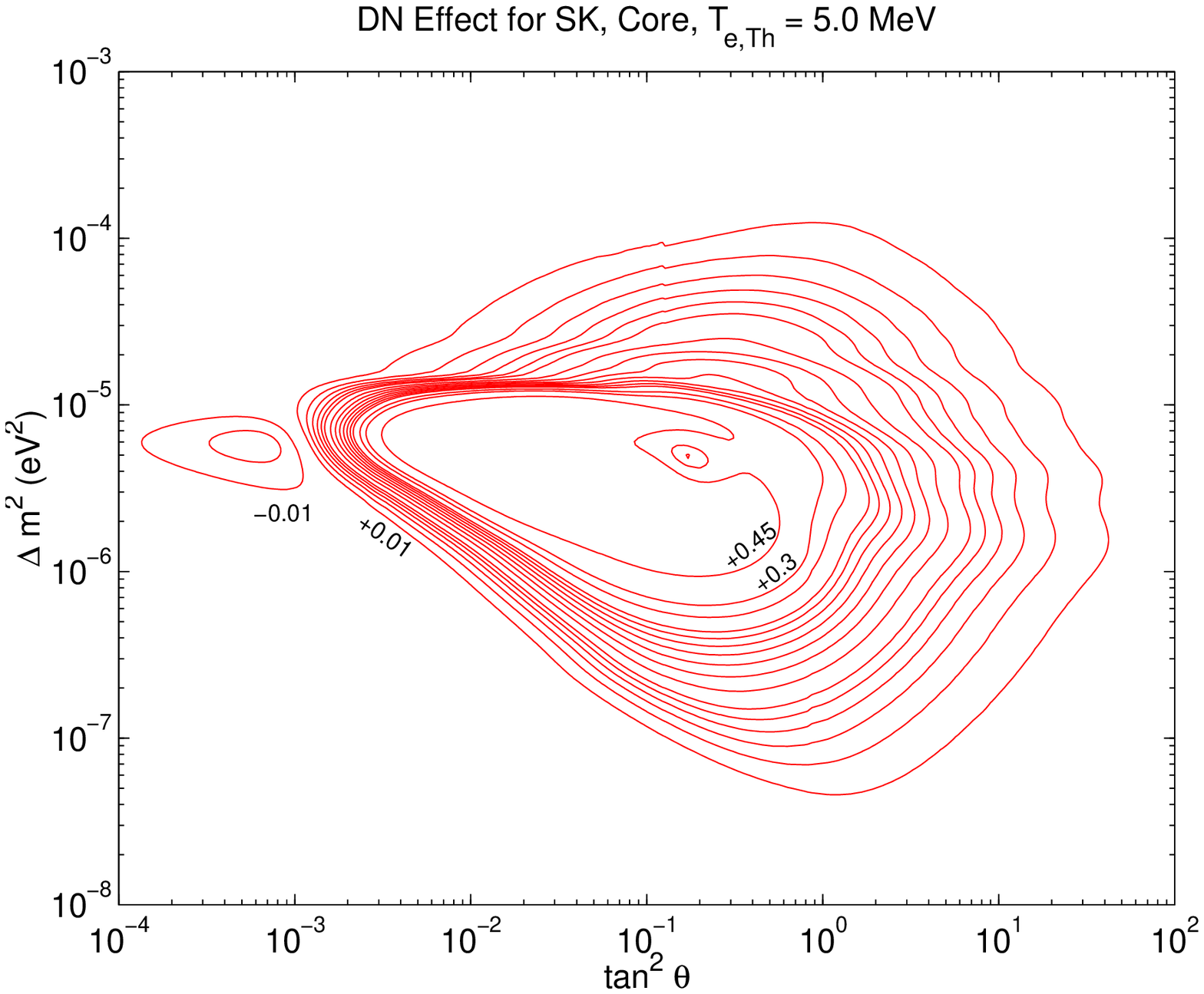,width=12.5cm}}\\
\vspace*{0.14truein}     %ORIGINAL SIZE=1.6TRUEIN x 100% - 0.2TRUEIN
   \caption{
Iso-(D-N) asymmetry
contour plot for the 
Super-Kamiokande experiment
for $T_{\mathrm{e},\mathrm{th}} = 5$\ MeV,
$T_{\mathrm{e},\mathrm{th}}$ being
the detected $e^{-}$ kinetic energy 
threshold. The contours correspond to values of 
the {\it Full Night} (upper panel) and 
{\it Core} (lower panel) asymmetries 
$A^{N,C}_{D-N}(SK) =-0.02,~-0.01,~ 0.01$, 0.02, 0.03,
% 0.04, 0.05, 0.06, 0.07,  0.08, 0.09, 0.10.
0.04, 0.05,
0.06, 0.08, 0.10, 0.12, 0.14, 0.16, 0.18, 0.20, 0.25, 0.30, 0.45.
  }\label{fig:sk:dn}
\end{figure}
%%%%%%%%%%%%%%%%%%%%%%%%%%%%%%%%%%%

%%%%%%%%%%%%%%%%%%%%%%%%%%%%%%%%
\begin{figure}[t]
  \centering
\vspace*{13pt}
% \leftline{\hfill\vbox{\hrule width 5cm height0.001pt}\hfill}
         \mbox{\epsfig{figure=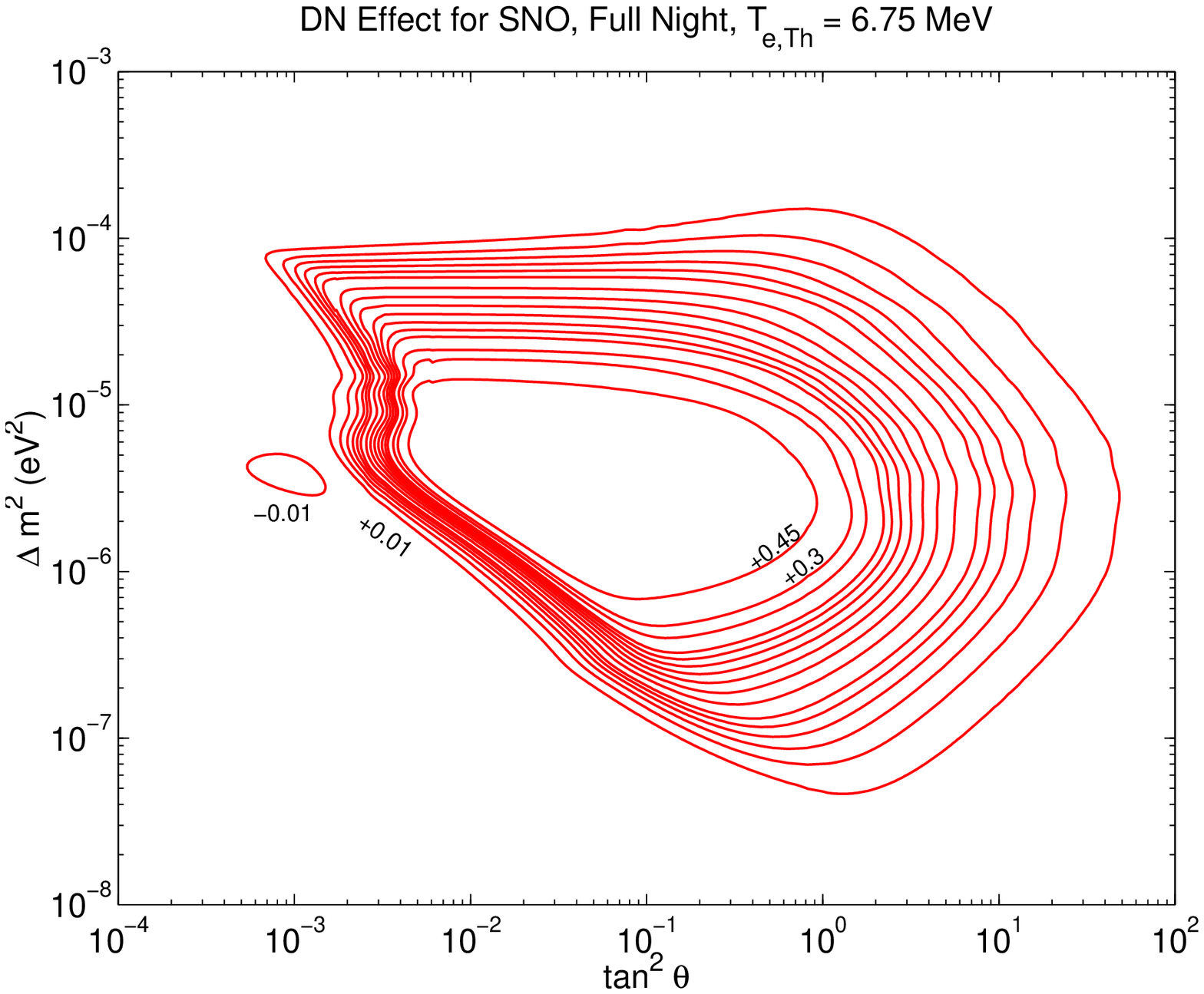,width=12.5cm}}\\
 \vspace{0.3cm} 
         \mbox{\epsfig{figure=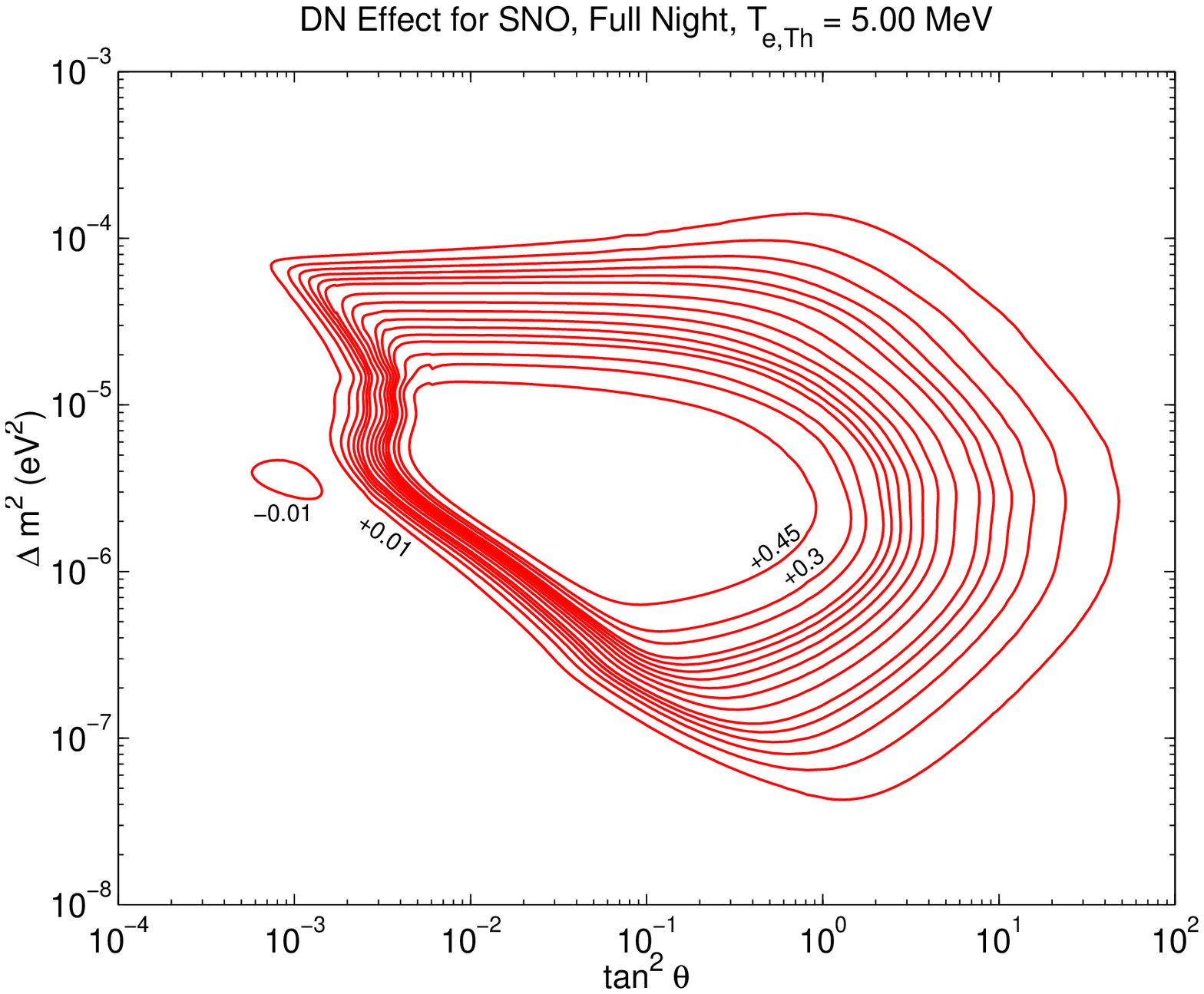,width=12.5cm}}
% \leftline{\hfill\vbox{\hrule width 5cm height0.001pt}\hfill}
\vspace*{0.045truein}     %ORIGINAL SIZE=1.6TRUEIN x 100% - 0.2TRUEIN
   \caption{
Iso-(D-N) asymmetry ($A^{N}_{D-N}(SNO)$) 
contour plot for the 
SNO experiment for
$T_{\mathrm{e},\mathrm{th}} = 6.75$\ MeV 
(upper panel) and
$T_{\mathrm{e},\mathrm{th}} = 5.00$\ MeV (lower
panel), $T_{\mathrm{e},\mathrm{th}}$ being
the (effective) kinetic energy 
threshold of the detected $e^{-}$ 
in the CC reaction. The contours 
correspond to values of 
the {\it Full Night} asymmetry
in the CC event rate 
$A^{N}_{D-N}(SNO) = -0.01,~0.01$, 0.02, 0.03, 0.04, 0.05,
0.06, 0.08, 0.10, 0.12, 0.14, 0.16, 0.18, 0.20, 0.25, 0.30, 0.45.
  }\label{fig:dn}
\end{figure}
%%%%%%%%%%%%%%%%%%%%%%%%%%%%%%%%%

%%%%%%%%%%%%%%%%%%%%%%%%%%%%%%%%
\begin{figure}[t]
  \centering
\vspace*{13pt}
% \leftline{\hfill\vbox{\hrule width 5cm height0.001pt}\hfill}
         \mbox{\epsfig{figure=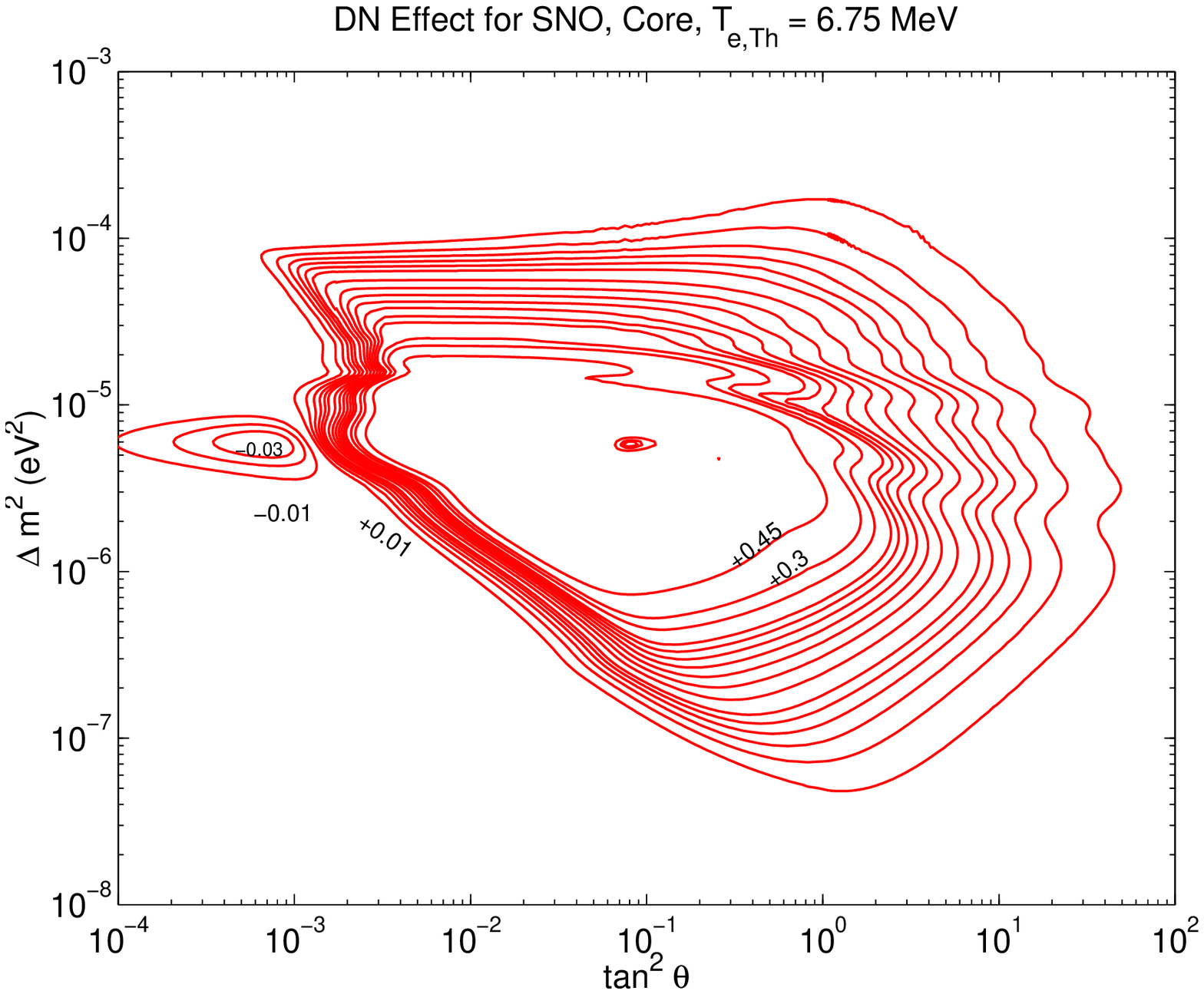,width=12.5cm}}\\
 \vspace{1cm}
         \mbox{\epsfig{figure=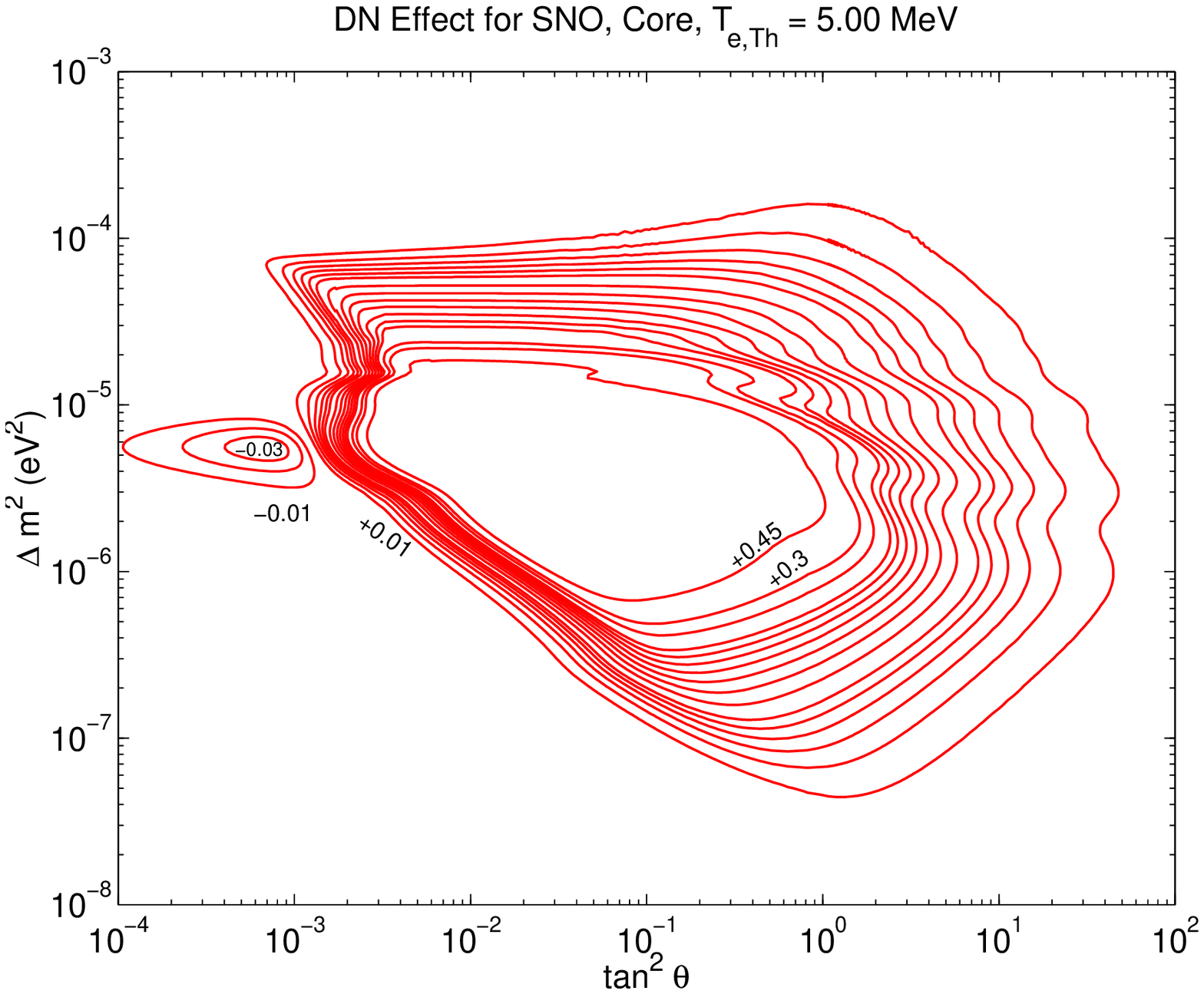,width=12.5cm}}
% \leftline{\hfill\vbox{\hrule width 5cm height0.001pt}\hfill}
\vspace*{0.045truein}     %ORIGINAL SIZE=1.6TRUEIN x 100% - 0.2TRUEIN
   \caption{
Iso-(D-N) asymmetry ($A^{C}_{D-N}(SNO)$) 
contour plot for the 
SNO experiment for
$T_{\mathrm{e},\mathrm{th}} = 6.75$\ MeV 
(upper panel) and
$T_{\mathrm{e},\mathrm{th}} = 5.00$\ MeV (lower
panel), $T_{\mathrm{e},\mathrm{th}}$ being
the (effective) kinetic energy 
threshold of the detected $e^{-}$ 
in the CC reaction. The contours 
correspond to values of 
the {\it Core} asymmetry
in the CC event rate 
$A^{C}_{D-N}(SNO) = -0.03,~-0.02,~-0.01,~0.01$, 0.02, 0.03, 0.04, 0.05,
0.06, 0.08, 0.10, 0.12, 0.14, 0.16, 0.18, 0.20, 0.25, 0.30, 0.45.
  }\label{fig:dn}
\end{figure}
%%%%%%%%%%%%%%%%%%%%%%%%%%%%%%%%%

%%%%%%%%%%%%%%%%%%%%%%%%%%%%%%%%%%%%%%%%%%
\begin{figure}[t]
 \centering  
\vspace*{13pt}
% \leftline{\hfill\vbox{\hrule width 5cm height0.001pt}\hfill}
         \mbox{\epsfig{figure=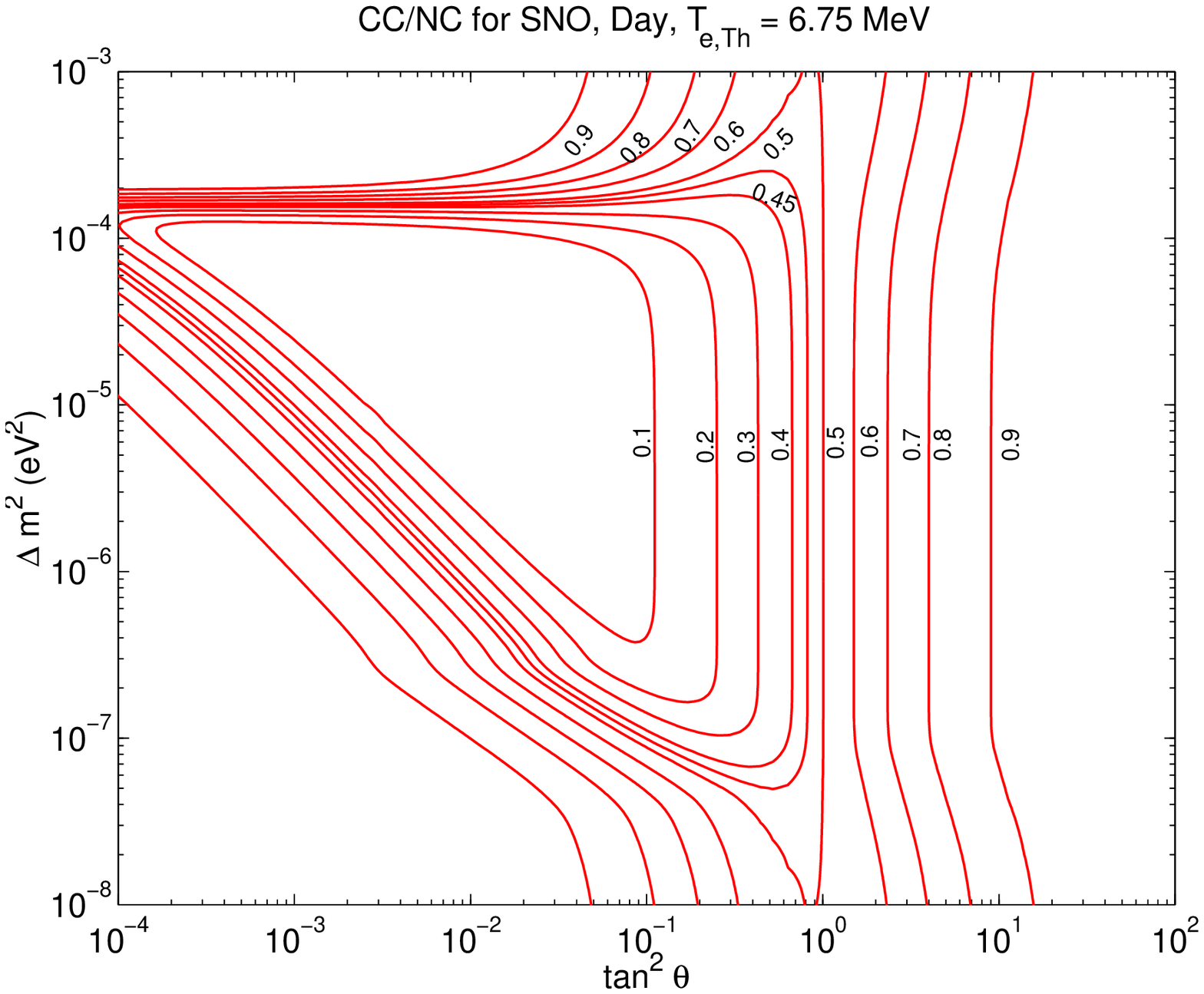,width=12.5cm}}\\
 \vspace{1.cm}
         \mbox{\epsfig{figure=red_QNo_T675_0_n5_tgts_mtx.eps,width=12.5cm}}
% \leftline{\hfill\vbox{\hrule width 5cm height0.001pt}\hfill}
\vspace*{0.045truein}     %ORIGINAL SIZE=1.6TRUEIN x 100% - 0.2TRUEIN
 \caption{
Iso$-R^{SNO}_{CC/NC}(D)$ contour plots 
for $T_{\mathrm{e},\mathrm{th}} = 6.75$\ MeV (upper panel) and
$T_{\mathrm{e},\mathrm{th}} = 5.00$\ MeV (lower
panel): each contour corresponds to a fixed value
of the {\it Day} ratio of the CC and NC solar neutrino 
event rates measured in the SNO experiment. In the case of 
absence of oscillations of solar neutrinos
$R^{SNO}_{CC/NC}(D) = 1$. The contours shown are for 
$R^{SNO}_{CC/NC}(D) = 0.10$, 0.20, 0.30, 0.40, 0.45, 0.50,
0.60, 0.70, 0.80, 0.90.
 }\label{fig:cc:nc}
 \end{figure}
%%%%%%%%%%%%%%%%%%%%%%%%%%%%%%%%%%%%%%%%%%%%%

%%%%%%%%%%%%%%%%%%%%%%%%%%%%%%%%%%%%%%%%%%
\begin{figure}[t]
 \centering
\vspace*{13pt}
% \leftline{\hfill\vbox{\hrule width 5cm height0.001pt}\hfill}
         \mbox{\epsfig{figure=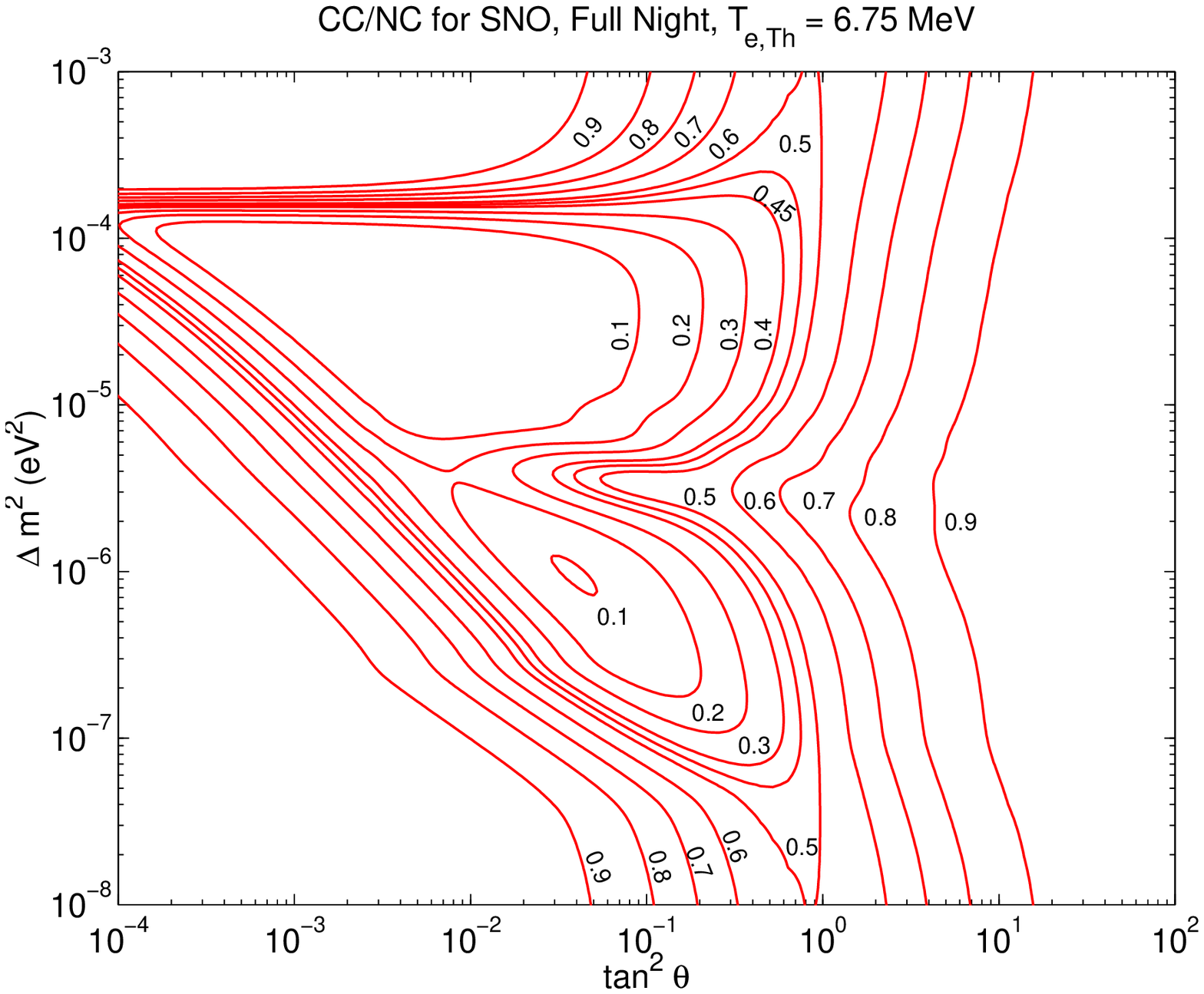,width=12.5cm}}\\
 \vspace{1.cm}
         \mbox{\epsfig{figure=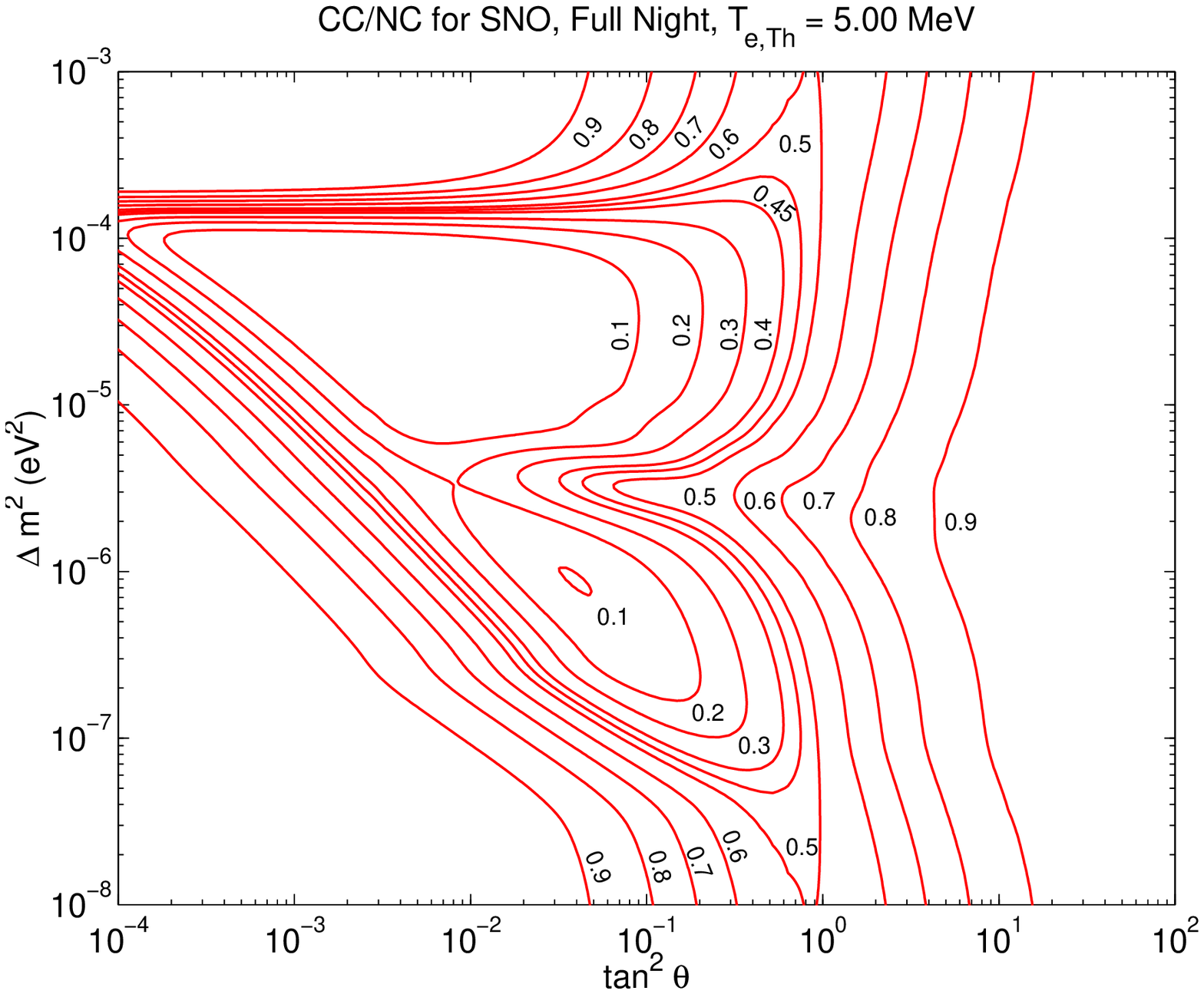,width=12.5cm}}
% \leftline{\hfill\vbox{\hrule width 5cm height0.001pt}\hfill}
\vspace*{0.045truein}     %ORIGINAL SIZE=1.6TRUEIN x 100% - 0.2TRUEIN
 \caption{
The same as in Fig. 4 for the {\it Night} 
ratio of the CC and NC solar neutrino 
event rates, $R^{SNO}_{CC/NC}(N)$,
measured in the SNO experiment. 
% The upper and lower panels correspond respectively 
% to $T_{\mathrm{e},\mathrm{th}} = 6.75$\ MeV and
% $T_{\mathrm{e},\mathrm{th}} = 5.00$\ MeV. 
In the case of absence of oscillations of solar neutrinos
$R^{SNO}_{CC/NC}(N) = 1$. 
The contours shown are for 
$R^{SNO}_{CC/NC}(N) = 0.10$, 0.20, 0.30, 0.40, 0.45, 0.50,
0.60, 0.70, 0.80, 0.90.
% Iso$-R^{SNO}_{CC/NC}(N)$ contour plots 
% for $T_{\mathrm{e},\mathrm{th}} = 5$\ MeV (upper panel) and
% $T_{\mathrm{e},\mathrm{th}} = 6.75$\ MeV (lower
% panel): each contour corresponds to a fixed value
% of the ratio of the CC and NC solar neutrino 
% event rates measured in the SNO experiment. 
 }\label{fig:cc:nc}
 \end{figure}
%%%%%%%%%%%%%%%%%%%%%%%%%%%%%%%%%%%%%%%%%%%%%

%%%%%%%%%%%%%%%%%%%%%%%%%%%%%%%%%%%%%%%%%%
\begin{figure}[t]
 \centering
\vspace*{13pt}
% \leftline{\hfill\vbox{\hrule width 5cm height0.001pt}\hfill}
         \mbox{\epsfig{figure=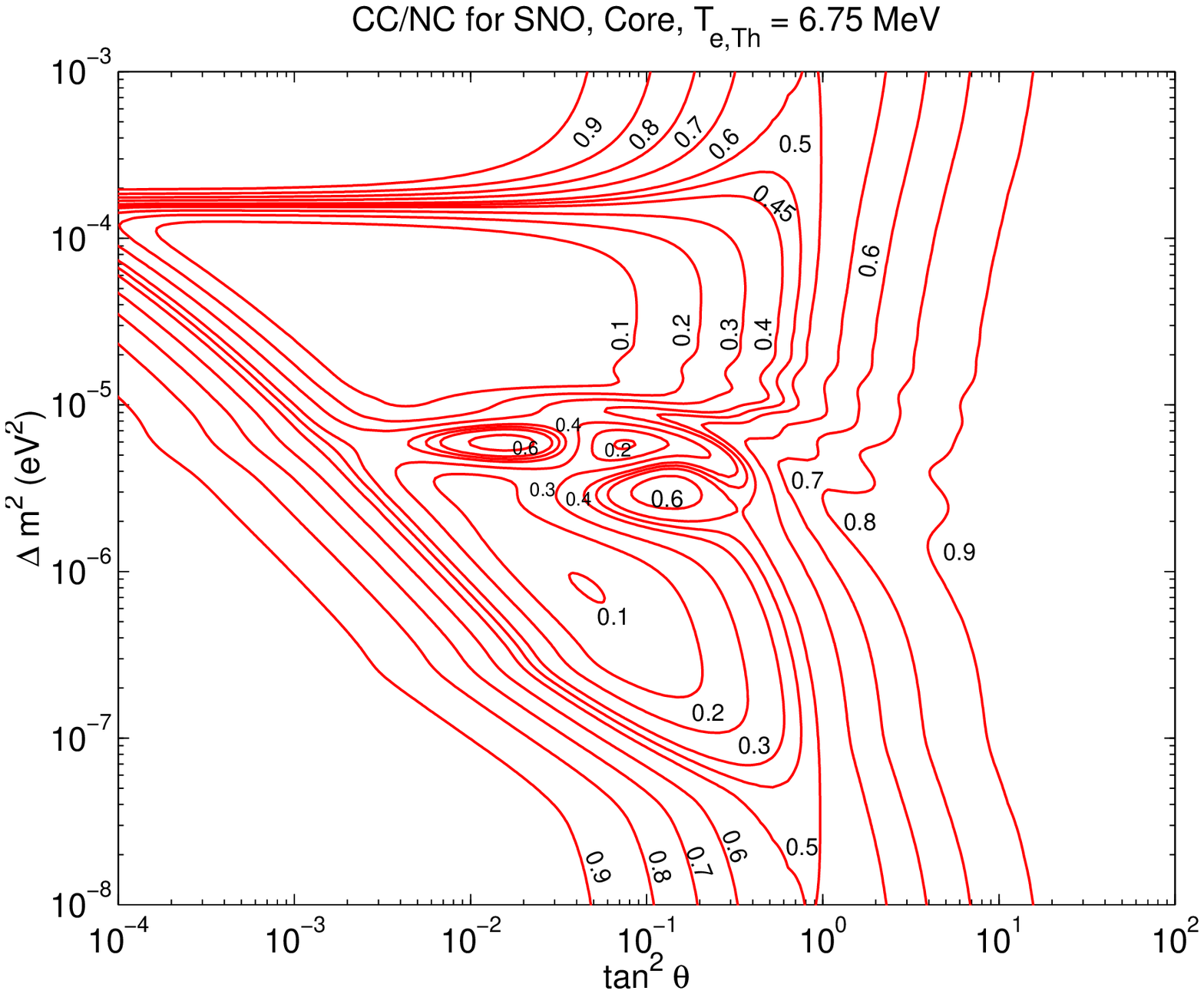,width=12.5cm}}\\
 \vspace{1.cm}
         \mbox{\epsfig{figure=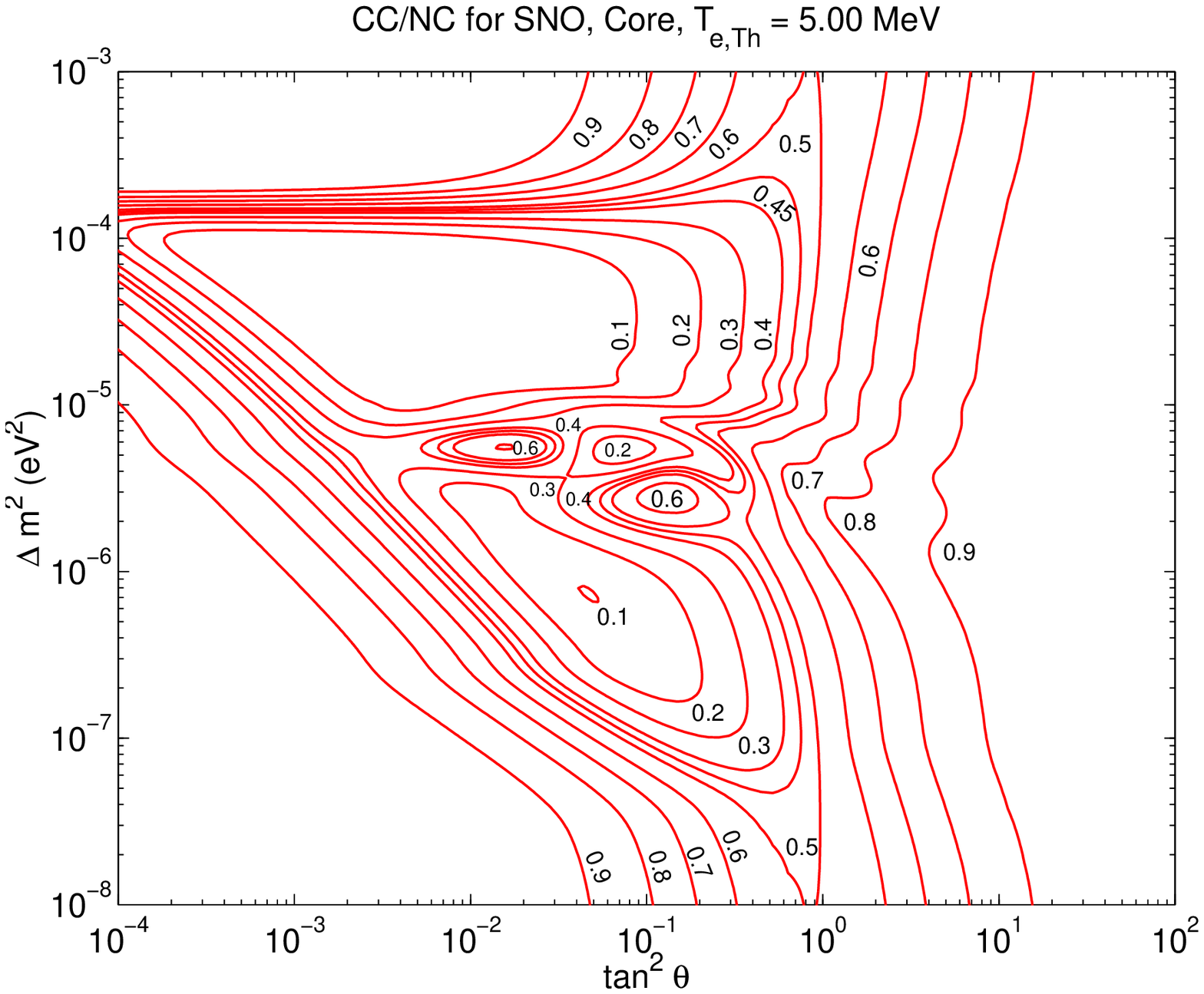,width=12.5cm}}
% \leftline{\hfill\vbox{\hrule width 5cm height0.001pt}\hfill}
\vspace*{0.045truein}     %ORIGINAL SIZE=1.6TRUEIN x 100% - 0.2TRUEIN
 \caption{
 The same as in Fig. 4 for the {\it Core} 
ratio of the CC and NC solar neutrino 
event rates, $R^{SNO}_{CC/NC}(C)$,
measured in the SNO experiment.
% , for $T_{\mathrm{e},\mathrm{th}} = 6.75$\ MeV 
% (upper panel)
% and $T_{\mathrm{e},\mathrm{th}} = 5.00$\ MeV
% (lower panel). 
In the case of absence of 
oscillations of solar neutrinos
$R^{SNO}_{CC/NC}(C) = 1$. 
The contours shown are for 
$R^{SNO}_{CC/NC}(C) = 0.10$, 0.20, 0.30, 0.40, 0.45, 0.50,
0.60, 0.70, 0.80, 0.90.
% Iso$-R^{SNO}_{CC/NC}$ contour plots 
% for $T_{\mathrm{e},\mathrm{th}} = 5$\ MeV (upper panel) and
% $T_{\mathrm{e},\mathrm{th}} = 6.75$\ MeV (lower
% panel): each contour corresponds to a fixed value
% of the ratio of the CC and NC solar neutrino 
% event rates measured in the SNO experiment. 
 }\label{fig:cc:nc}
 \end{figure}
%%%%%%%%%%%%%%%%%%%%%%%%%%%%%%%%%%%%%%%%%%%%%

\end{document}